\documentclass[aps,rmp,reprint,amsmath,amssymb,graphicx,longbibliography]{revtex4-1}
\usepackage[T1]{fontenc}
\usepackage{bm}
\usepackage{graphicx}
\usepackage{dcolumn}
\usepackage{appendix}
\usepackage{bm}
\usepackage[usenames,dvipsnames]{xcolor}
\usepackage{subfigure}
\usepackage{bbm}
\usepackage{enumerate}
\usepackage[
  bookmarks=true,
  colorlinks,
  linkcolor=blue,
  urlcolor=OliveGreen,
  citecolor=OliveGreen,
  plainpages=false,
  pdfpagelabels,
  final,
  breaklinks=true
]{hyperref}
\usepackage{dsfont}
\usepackage{physics}
\usepackage{amsthm}

\newcommand{\relaxket}[1]{\lvert{#1}\rangle}

\begin{document}

\title{\textit{Colloquium:} Quantum optics of intense light--matter interaction}

\author{Philipp Stammer}
\email{philipp.stammer@icfo.eu}
\affiliation{ICFO -- Institut de Ciencies Fotoniques, The Barcelona Institute of Science and Technology, Castelldefels (Barcelona) 08860, Spain.}
\affiliation{Atominstitut, Technische Universit\"{a}t Wien, 1020 Vienna, Austria}

\author{Javier Rivera-Dean}
\email{physics.jriveradean@proton.me}
\affiliation{ICFO -- Institut de Ciencies Fotoniques, The Barcelona Institute of Science and Technology, Castelldefels (Barcelona) 08860, Spain.}

\author{Paraskevas Tzallas}
\email{ptzallas@iesl.forth.gr}
\affiliation{Foundation for Research and Technology-Hellas, Institute of Electronic Structure \& Laser, GR-70013 Heraklion (Crete), Greece}
\affiliation{Center for Quantum Science and Technologies (FORTH-QuTech), GR-70013 Heraklion (Crete), Greece.}
\affiliation{ELI-ALPS, ELI-Hu Non-Profit Ltd., Dugonics tér 13, H-6720 Szeged, Hungary}

\author{Marcelo F. Ciappina}
\email{marcelo.ciappina@gtiit.edu.cn}
\affiliation{Department of Physics, Guangdong Technion - Israel Institute of Technology, 241 Daxue Road, Shantou, Guangdong, China, 515063}
\affiliation{Technion - Israel Institute of Technology, Haifa, 32000, Israel}
\affiliation{Guangdong Provincial Key Laboratory of Materials and Technologies for Energy Conversion, Guangdong Technion - Israel Institute of Technology, 241 Daxue Road, Shantou, Guangdong, China, 515063}

\author{Maciej Lewenstein}
\email{maciej.lewenstein@icfo.eu}
\affiliation{ICFO -- Institut de Ciencies Fotoniques, The Barcelona Institute of Science and Technology, Castelldefels (Barcelona) 08860, Spain.}
\affiliation{ICREA, Pg. Llu\'{\i}s Companys 23, 08010 Barcelona, Spain}

\date{\today{}}

\begin{abstract}
Intense light–matter interaction largely relies on the use of high-power light sources, creating fields comparable to, or even stronger than, the field keeping the electrons bound in atoms. Under such conditions, the interaction induces highly nonlinear processes such as high harmonic generation, in which the low-frequency photons of a driving laser field are upconverted into higher-frequency photons. These processes have enabled numerous groundbreaking advances in atomic, molecular, and optical physics, and they form the foundation of attosecond science. Until recently, however, such processes were typically described using semi-classical approximations, since the quantum properties of the light field were not required to explain the observables. This has changed in the recent past. Ongoing theoretical and experimental advances show that fully quantized descriptions of intense light–matter interactions, which explicitly incorporate the quantum nature of the light field, open new avenues for both fundamental research and technological applications at the fully quantized level.~These advances emerge from the convergence of quantum optics with strong-field physics and ultrafast science.~Together, they have given rise to the field of quantum optics and quantum electrodynamics of strong-field processes.

\end{abstract}

\maketitle

\tableofcontents{}

\section{Introduction}
\label{intro}

\begin{figure}
    \centering
    \includegraphics[width=1.0\columnwidth]{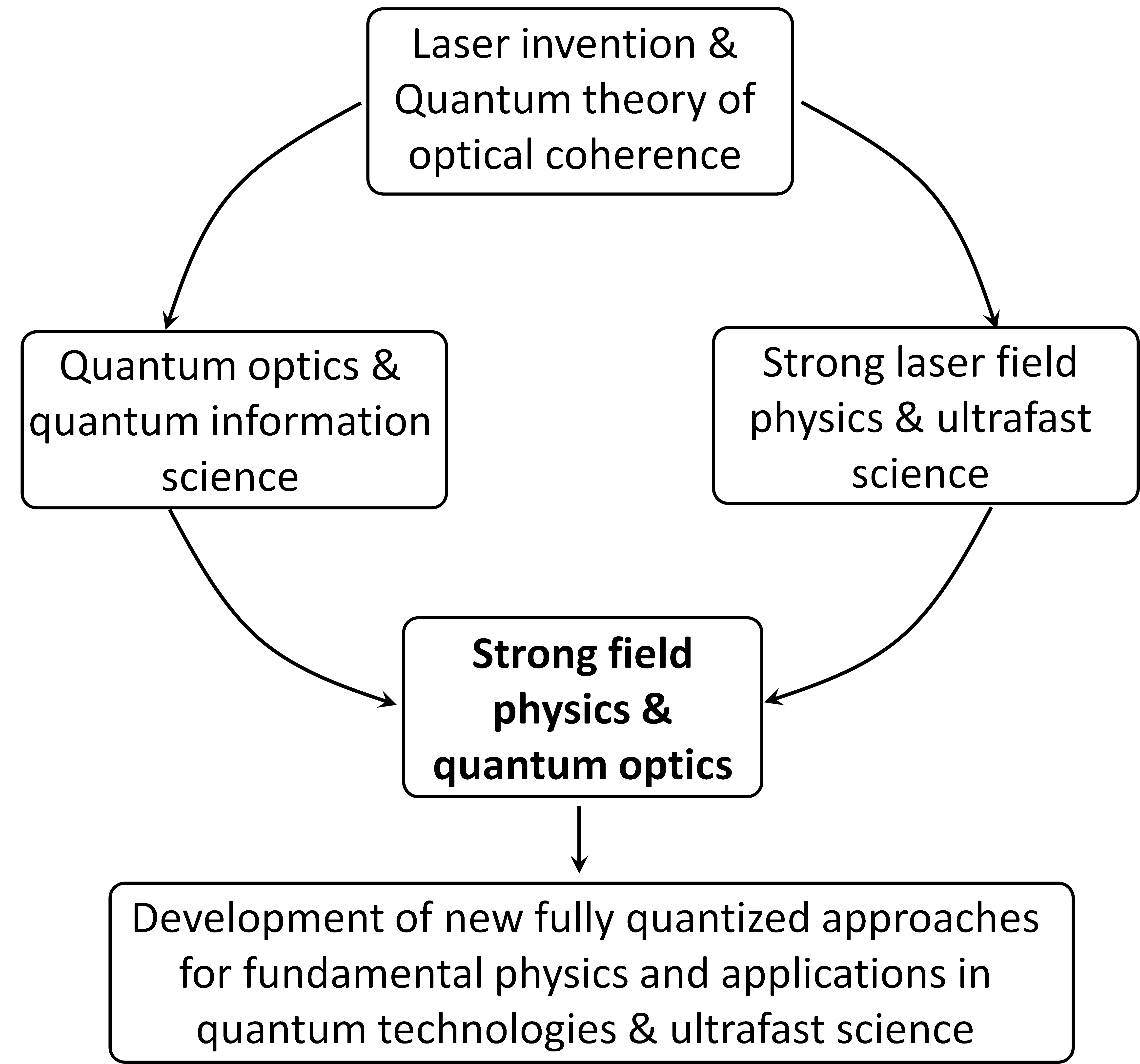}
    \caption{With the advent of the laser and the theory of quantum optics and optical coherence, two important but distinct fields have emerged. On one side lies the low-photon-number regime, where quantum optical descriptions of photons have led to remarkable applications in quantum information science; on the other side is the high-photon-number regime of strong laser-field physics, with applications in ultrafast science. Recent progress has demonstrated how these previously separate fields can now be integrated.}
    \label{fig:advances}
\end{figure}

After the invention of the laser \cite{Maiman_Nature_1960} and the development of the quantum theory of optical coherence~\cite{glauber1963quantum, glauber1963coherent}, research on light-matter interaction largely followed two distinct directions [Fig.~\ref{fig:advances}]. On one side, the field of Quantum Optics (QO), which relies on the use of relatively low photon number light sources and the development of fully quantized theories to describe light–matter interactions~\cite{cohen2024atom, cohen2024photons}. On the other side, the field of Strong Field Physics (SFP) which relies on the use of high-power lasers and semiclassical approaches to describe the laser-matter interaction~\cite{KrauszIvanov}.
Both directions have led to groundbreaking achievements, ranging from Nobel Prize-winning experiments to advances across interdisciplinary areas in basic research and technology. However, despite this remarkable progress, the two fields have mostly remained disconnected, leaving unexplored the potential advantages arising from their integration. 
In this Colloquium we present the convergence of SFP with QO and the advantages, in basic research and technologies, emerging from this connection.

\subsection{Motivation to connect quantum optics and strong field physics}

The synergy between QO and SFP aims to combine tools and methods from these domains to create new techniques, and light sources for fundamental research, as well as for novel applications in quantum technology and ultrafast science at a fully quantized level. In this direction, tremendous theoretical and experimental progress has been achieved in recent years~\cite{stammer2023quantum, bhattacharya2023strong, lamprou2024generation, cruz2024quantum}. The field has begun to evolve into its own area of research, often referred to as \textit{Extreme Quantum Optics}, which explores quantum optics and quantum electrodynamics in strong-field regimes, traditionally considered as classical. For completeness, and to emphasize the significance of \textit{Extreme Quantum Optics}, it is useful in this introductory section to briefly discuss some of the fundamental concepts and breakthrough achievements in the two distinct areas of QO and SFP, and to describe how these can be integrated within a unified framework for novel applications in fundamental research and technology.

In QO, the majority of the experiments and applications are conducted using relatively low photon number light sources. The light-matter interaction is described by fully quantized approaches~\cite{scully1997quantum, cohen2024photons, cohen2024atom}, which are essential for  characterizing the joint properties of the light-matter system. This includes developing methods for characterizing and controlling its quantum state~\cite{BSM97, LR09, mandel1995optical, Bachor_book_2019, Blume2010, Leonhardt_book_1997}, methods that largely rely on statistical measurements governed by quantum noise, as well as designing techniques for engineering entangled states and protocols for their classification and characterization~\cite{Huber_EntanglementCertif_NatRevPhys_2019}. Today, quantum optics is at the core of modern quantum technologies~\cite{Acin2018, Walmsay2015, dowling2003quantum, o2009photonic}, ranging from applications in quantum computation, communication, and sensing to fundamental tests of quantum theory~\cite{Haroche2013, Wineland2013, Gilchrist2004, Giovannetti2004, Giovannetti2011, Jouguet2013, Lloyd1999, Ralph2003, Joo2011, Schnabel2017}. It provides methods for engineering non-classical (quantum) light states, such as Fock states, squeezed states, and coherent state superpositions (optical Schrödinger cat states)~\cite{BSM97, Leonhardt_book_1997, Andersen2016, Kimble1977, Diedrich1987, Haroche2013, Wineland2013, Zavatta2004, Ourjoumtsev2006, Hacker2019, Sychev2017}, as well as entangled light states, all central to the optical quantum technology platforms.

On the other side, in SFP, the majority of the experiments and applications are conducted using intense light sources typically obtained through high power femtosecond laser sources~\cite{Mourou2019, Strickland2019}. These lasers can provide electric fields with strength comparable to, or even stronger than, the field keeping the electrons bound in atoms~\cite{protopapas_atomic_1997, Krause1992PRL, corkum_plasma_1993, Maciej3Step}. In this case, the light-matter interaction is usually described by semiclassical approximations where the electromagnetic radiation is treated classically and matter quantum mechanically~\cite{Maciej3Step, amini2019symphony}. A central aspect in SFP involves accessing the ultrafast electron dynamics of the system in the presence of a strong light field. This includes developing methods for controlling the wavefront of the driving laser field and tracing ultrafast electron dynamics, as well as designing techniques for engineering ultrashort pulses and establishing protocols for temporal pulse characterization~\cite{Huillier_Nobel_2024, Krausz_Nobel_2024, Agostini_Nobel_2024}. Strong field physics lies at the core of numerous remarkable achievements in atomic, molecular, and optical physics, ultrashort pulse engineering, and ultrafast science~\cite{KrauszIvanov}, with applications ranging from relativistic particle acceleration and laser--plasma interaction~\cite{Mourou2019}, to high harmonic generation (HHG)~\cite{ferray-multiple-1988, Li1989}, high-resolution spectroscopy~\cite{Gohle2005, Cingoz2012}, and attosecond science~\cite{KrauszIvanov}. The majority of these achievements, owing to the high photon number of the light fields, have been sufficiently explained without considering the quantum nature of light. Therefore, the quantum aspects of the light field have largely been ignored, and the advantages arising from the connection between QO and SFP have remained unexplored.

In this Colloquium, we describe how the field of \textit{Extreme Quantum Optics}, can provide unique opportunities to overcome limitations of the current technologies and develop new approaches for fundamental research and technology at a fully quantized level.

The structure of this Colloquium is as follows:
In Sec.~\ref{sec:semiclassical} and \ref{sec:old_QO} we provide a brief recapitulation of the semi-classical picture of strong field phenomena and a short summary of the very early efforts of quantum optical attempts towards those phenomena, respectively. In Sec.~\ref{Sec:Quant:state}-\ref{Sec:Entanglement} we introduce the quantum optical approach towards HHG, discussing the solution of the quantum state of the harmonic radiation including non-classical effects. In Sec.~\ref{sec:coherence} we discuss the quantum optical theory of optical coherence for the process of HHG, followed by a discussion on complex systems in Sec.~\ref{sec:complex}. In Sec.~\ref{sec:conditioning} we discuss conditioning schemes in HHG, leading to the generation of high-photon-number Schrödinger cat states, while Sec.~\ref{sec:beyond_coherent} and \ref{sec:propagation} discuss non-classical driving fields for HHG. 
After the process of HHG, we discuss quantum optical strong-field ionization in Sec.~\ref{sec:ATI_part1} and \ref{sec:ATI_quantum}, when classical and quantum light drives this process, respectively. 
In Sec.~\ref{sec:application} we present the road towards applications in quantum technologies, including analog simulation of strong field processes in Sec.~\ref{sec:analog}. We conclude this Colloquium with a discussion in Sec.~\ref{sec:discussion} and outlook for future directions in Sec.~\ref{sec:outlook}.

\subsection{The semiclassical approach}
\label{sec:semiclassical}

\begin{figure}
    \centering
    \includegraphics[width=1.0\columnwidth]{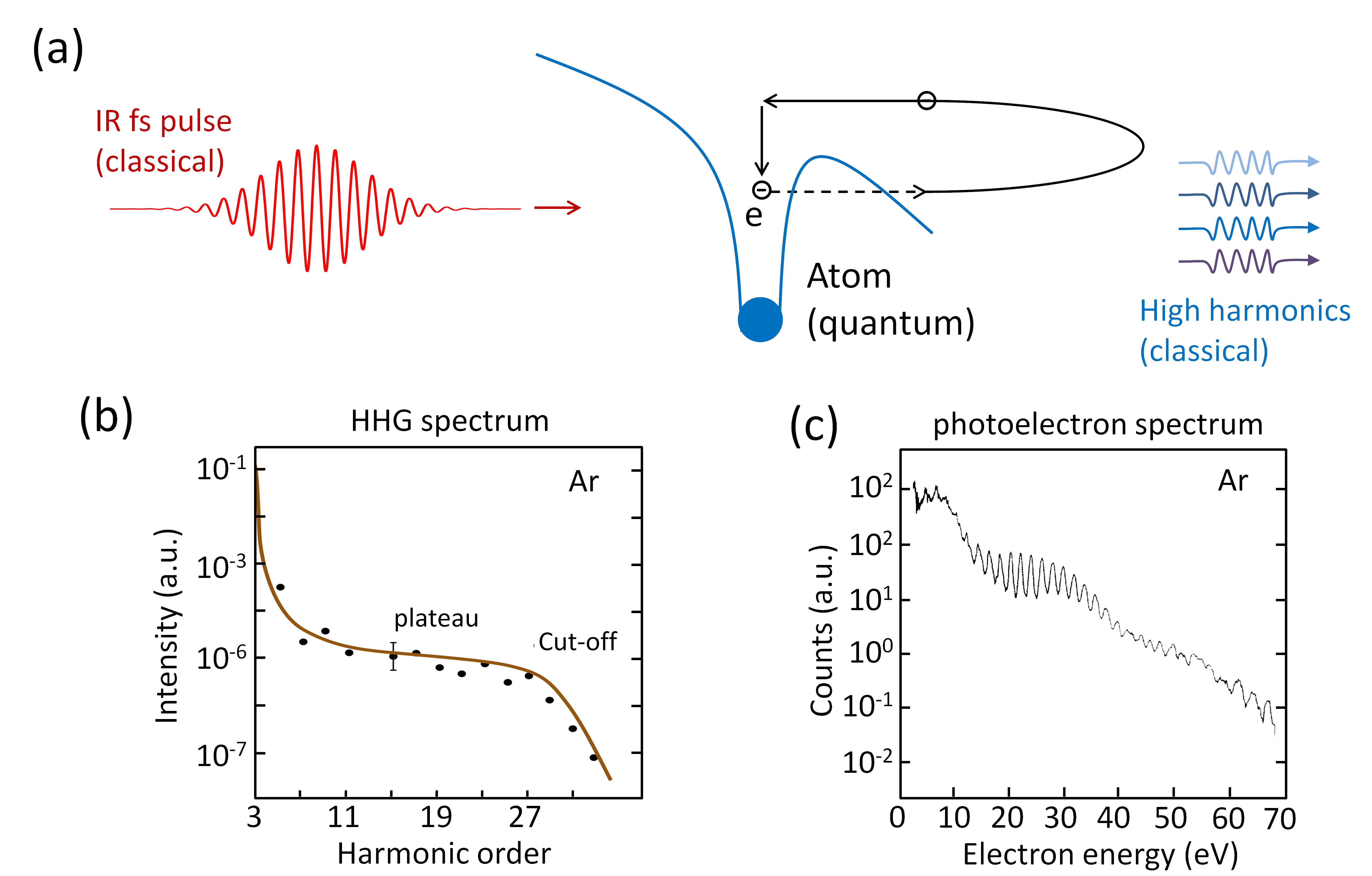}
    \caption{A schematic of the semiclassical approach to the interaction of an atom with an intense infrared (IR) laser pulse of femtosecond (fs) duration. The light field is treated classically and the atom quantum mechanically. (a) Electron recollision picture. (b) Intensity of high harmonics (showing the plateau and cut-off regions) generated by the interaction of an intense linearly polarized laser pulse with Argon atoms.  (c) Photoelectron spectrum generated by the interaction of Argon atoms with an intense linearly polarized laser pulse. Figs.~(b) and (c) reproduced from \cite{Huillier_Nobel_2024} and \cite{Paulus1994PRL}, respectively.}
    \label{fig:Fig_SemiClass_RMP}
\end{figure}

The semiclassical approach has been a cornerstone in understanding strong-field physics phenomena. Its development was motivated by the failure of perturbative nonlinear optics at intensities where the electric field of the driving laser becomes comparable to the Coulomb field binding the electron to the nucleus. In this regime, multiphoton ionization and above-threshold ionization (ATI) spectra revealed signatures that could not be explained within standard perturbation theory~\cite{agostini_free-free_1979,Eberly1989JOSAB,Eberly1989PRL}. A conceptual breakthrough came with the introduction of the so-called three-step model, which provides a simple yet powerful picture: (i) an electron tunnels through the Coulomb barrier suppressed by the strong classical field, (ii) the liberated electron is accelerated by the laser field, and (iii) it may revisit its parent ion, leading either to rescattering or to recombination accompanied by the emission of high-order harmonics~\cite{corkum_plasma_1993,Schafer1993PRL,Maciej3Step} [Fig.~\ref{fig:Fig_SemiClass_RMP} (a)].

Typical HHG spectra in atoms, molecules and solids are characterized by a rapid falloff at low harmonic orders, followed by a broad plateau of nearly constant intensity that extends over many harmonics, culminating in a sharp cutoff that defines the maximum photon energy. The recombination step of the model explains the characteristic plateau in the harmonic spectrum [Fig.~\ref{fig:Fig_SemiClass_RMP} (b)], extending up to the cutoff law\footnote{Here, $U_p=e^2E_0^2/4m\omega_L^2$ is the ponderomotive energy, with $E_0$ the peak electric field amplitude and $\omega_L$ is the driving laser frequency and $I_p$ is the ionization potential.} $\hbar \omega_{\mathrm{cutoff}} \simeq 3.17 U_p+I_p$, a result derived both numerically and semiclassically~\cite{Krause1992PRL,LHuillier1993PRL, Maciej3Step}. This relation has been verified in numerous experiments, establishing HHG as a robust tool for producing coherent extreme-ultraviolet (XUV) radiation and attosecond pulses~\cite{Agostini2004RPP}. On the ATI side, the semiclassical approach faithfully predicts both the energy limits in the direct ($2U_p$) and rescattered electrons ($10U_p$), in remarkable agreement with experimental observations. The $2U_p$ cutoff corresponds to electrons that are released near a field maximum and subsequently accelerated away without revisiting the parent ion, while the $10U_p$ cutoff arises from those electrons that are driven back to the ionic core and elastically rescatter, gaining substantially more energy from the laser field. These two limits not only delineate the structure of ATI spectra [Fig.~\ref{fig:Fig_SemiClass_RMP} (c)] but also serve as clear signatures of the underlying rescattering mechanism that links ATI and HHG within the same semiclassical framework~\cite{Paulus1994PRL}.

The semiclassical approach is not only predictive but also intuitive. The rescattering mechanism explains the plateau in ATI spectra and the universality of HHG across atomic species~\cite{Eberly1989PRL}, while also providing a natural interpretation of correlated electron dynamics in terms of trajectories driven back to the core. Building on this intuition, the \emph{strong-field approximation} (SFA) was developed as a quantitative framework. Originally formulated in the KFR theory~\cite{Keldysh1965,Faisal1973,Reiss1980}, SFA assumes that after ionization the electron propagates as a Volkov state (a free electron dressed by the laser field), neglecting the Coulomb interaction during the continuum motion. The atomic potential enters only at ionization and recombination (for HHG) or rescattering (for ATI). Despite its simplifications, the SFA captures the essential physics of tunneling, propagation, and recollision, and has been refined over the decades into a widely used theoretical tool for analyzing strong-field phenomena~\cite{Maciej3Step,amini2019symphony}.

The starting point of the SFA is the semiclassical Hamiltonian, that can be written as
\begin{align}
\label{eq:hamiltonian_semi}
    H_{sc}(t) =H_S + H_{I,sc}(t),
\end{align}
where $H_S$ is the system Hamiltonian of a single electron
\begin{align}
\label{eq:hamiltonian_atom}
    H_S = -\frac{\hbar^2}{2m}\nabla^2 + V(\mathbf{r}),
\end{align}
bound by the potential $V(\mathbf{r})$. In Eq.~(\ref{eq:hamiltonian_semi}), $H_{I,sc} (t)$ is the dipole coupling which describes the interaction of the atomic or molecular system with the laser radiation. In the length gauge we have $H_{I,sc}(t)=-\mathbf{d}\cdot\mathbf{E}_{cl}(t)$, where $\mathbf{E}_{cl}(t)$ is the classical laser electric field and $\mathbf{d}=e\mathbf{r}$ the dipole moment, with $e$ being the electron charge.

Within the SFA, the time-dependent dipole moment expectation value $\expval{\mathbf{d}(t)}$ responsible for HHG is~\cite{Maciej3Step}
\begin{eqnarray}
\label{tdipole}
\expval{\mathbf{d}(t)} & =&  -i\frac{e}{\hbar} \int_{-\infty}^{t} dt' \int d\mathbf{p}\;
\mathbf{d}^{*}\!\left[\mathbf{p}-e\mathbf{A}_{cl}(t)\right]\,
e^{-iS(\mathbf{p},t,t')/\hbar} \nonumber\\
& & \times \mathbf{E}_{cl}(t') \cdot  \mathbf{d}\!\left[\mathbf{p}-e\mathbf{A}_{cl}(t')\right] + \text{c.c.},
\end{eqnarray}
where the semi-classical action is
\begin{equation}
S(\mathbf{p},t,t') = \int_{t'}^{t} d\tau \left( \frac{[\mathbf{p}-e\mathbf{A}_{cl}(\tau)]^2}{2m} + I_p \right).
\end{equation}
Here $t'$ is the ionization time, $\mathbf{p}$ the canonical momentum in the continuum, $\mathbf{A}_{cl}(t)$ the classical laser vector potential and $\mathbf{E}_{cl}(t')$ the instantaneous classical laser electric field at ionization. In Eq.~(\ref{tdipole}), $\mathbf{d}(\mathbf{v})$, with kinetic momentum $m\mathbf{v}=\mathbf{p}-e\mathbf{A}_{cl}(t)$, is the bound-free transition dipole matrix element
\begin{align}
\label{dipole}
    \mathbf{d}(\mathbf{v}) = e\langle\mathbf{v}|\mathbf{r}|g\rangle,
\end{align}
where $|\mathbf{v}\rangle$ ($|g\rangle$) is the free (ground) electronic state. The harmonic spectrum can then be written as
\begin{align}
\label{FTdipole}
    I(\omega)\propto\omega^{4}|\tilde{\mathbf{d}}(\omega)|^{2},
\end{align}
where $\tilde{\mathbf{d}}(\omega)= FT [\expval{\mathbf{d}(t)}]$ is the Fourier transform of $\expval{\mathbf{d}(t)}$, and naturally leads to the $3.17U_p$-cutoff law.

While the semiclassical approach has been remarkably successful in capturing the main features of strong-field dynamics, it suffers from important limitations~\cite{stammer2024limitations}. Among them is the inability to account for the full quantum character of the emitted radiation, including entanglement, photon statistics, or quantum optical coherence. Ultimately, the semiclassical approach has shaped the modern understanding of attosecond science. By linking the generation of high harmonics with electron trajectories on sub-femtosecond timescales, it established the foundation for attosecond pulse synthesis~\cite{Agostini2004RPP}. Today, semiclassical theory continues to guide both fundamental studies of ultrafast electron dynamics and technological applications, from coherent XUV sources to petahertz electronics. Its ability to connect simple physical intuition with accurate predictions exemplifies its enduring value in strong-field and ultrafast physics.

\subsection{Past attempts on quantum optical description of intense light--matter interaction }
\label{sec:old_QO}

Despite the tremendous success of the semiclassical approach to strong field driven processes, several works have attempted a full quantum description of strong-field ionization, HHG in intense laser fields and even the macroscopic description of the laser-matter interaction within a full QED framework~\cite{sundaram1990high, bogatskaya2016polarization, diestler2008harmonic, gao2000interpretation, becker1997unified, compagno1994qed, xu1993non, Gauthey1995, eberly1992spectrum, ishkhanyan2021markoff, eden2000eden, chen2000comment, gao1998quantum, Wang2012, wang2007frequency, fu2001interrelation, gao2000nonperturbative, guo1992multiphoton, aaberg1991scattering, guo1989scattering, guo1988quantum, Sindelka2010, Yangaliev2020, Gombkoto2020, Varro2020, Hu2008, Kuchiev1999, Usachenko2002, Bogatskaya2017, Burenkov2010, gombkotHo2016quantum, gombkotHo2021quantum, gombkotHo2024parametric}. 
Nevertheless, while these studies have provided valuable insights, the success of these approaches hold off. On the one hand, these attempts have not directly identified observables that can be immediately associated with the quantum noise of the EM field, revealed signatures that can be clearly distinguished from the observables described by semiclassical methods, or relied on oversimplified models.

As a result, fundamental questions, such as the quantum state of the field after HHG, including the back-action of the interaction on the state of the field, the nature of the radiation emitted after the interaction, light-matter entanglement, and entanglement between field modes, have remained largely unexplored. Consequently, potential applications of intense light–matter interactions in quantum technology have largely been left open. 

However, it remains to acknowledge that some of the earlier papers have already identified crucial aspects on the origin of quantum optical signatures. Noteworthy is the work by Sundaram and Miloni~\cite{sundaram1990high}, which have shown that neglecting dipole moment correlations, and the number of emitters are crucial to reproduce the results known from classical theory.

\section{Quantum optical HHG revisited}\label{Sec:QHHG:revisited}

In recent years, it has been demonstrated that intense light-driven processes, particularly HHG and ATI, allow for the generation of high-photon-number non-classical states of light. This includes optical Schrödinger cat states \cite{lewenstein2021generation, rivera2022strong, stammer2022theory, stammer2022high, rivera2022light}, multi-mode squeezed fields \cite{Gorlach2020, stammer2024entanglement, Misha2025resonant, theidel2024evidence, lange_excitonic_2025}, and light-matter entangled states that can all be engineered using strong-field processes, and moves quantum optics firmly into the high-intensity regime. Moreover, bright squeezed states with intensities bright enough to drive or perturb HHG~\cite{spasibko2017multiphoton,manceau2019indefinite-mean, rasputnyi2024high, lemieux2024photon} have already been realized. These developments, together with numerous further articles, have revisited and expanded the perspective on the intersection of strong field physics and quantum optics~\cite{stammer2023quantum, bhattacharya2023strong, lamprou2024generation, cruz2024quantum}.

In this section, we revisit the quantum optical approach to the process of HHG, and provide an overview on the latest developments taking into account the quantum nature of the EM-field in strong laser driven processes~\cite{stammer2023quantum, cruz2024quantum}. We discuss theoretical predictions, which go beyond the semiclassical picture, and emphasize on experiments in which non-classical signatures are observed in the measurement statistics.

\subsection{The quantum state of the field}\label{Sec:Quant:state}

Quantization of the radiation field in intense laser driven processes naturally leads to the prompt questions about: \textit{(i) What is the quantum state of the field after the interaction? (ii) Are there immanent quantum signatures of the photon statistics in the process hidden in semi-classical models? And (iii) How can we use the quantum nature of the field for our advantage?} 
We shall answer these questions one by one, and discuss where the quantum optical approach deviates from the semi-classical picture. First, we will introduce the theoretical formalism necessary to understand the consequences of the fully quantized interaction, which allows to compute the quantum state of the field in the process of HHG. 

When considering the quantum nature of the process of HHG we need to include the operators acting on the Hilbert space of the field, as well as the initial quantum state taking into account the experimental boundary conditions of the driving laser. For the former we include the Hamiltonian of the field $H_F$, and the quantized light-matter coupling, while for the latter we consider (for now) driving the process by an intense laser field described by a coherent state $\ket{\alpha}$ of amplitude $\alpha = \abs{\alpha} e^{i \phi}$. The harmonics are initially in the vacuum $\ket{\{0_q\}} = \bigotimes_{q \ge 2} \ket{0_q}$.
We can see that the field is initially in a product of Gaussian coherent states, and we will see below that the dynamics under the strong field assumptions in HHG give rise to a product of Gaussian coherent states in the final field state~\cite{lewenstein2021generation}. 
This effectively describes the process of HHG as a mapping of coherent input states to coherent output states, and we shall see under which assumptions and approximations this is justified~\cite{cruz2024quantum}, while the non-linear dynamics of the laser-driven electron is encoded in the coherent state amplitudes of the generated harmonics. We already want to point out that this mapping is often implicitly assumed in works beyond coherent driving fields, such as squeezed states of light, as we will examine below in Sec.~\ref{sec:beyond_coherent}. We also discuss the consequences of these approximations in the subsequent sections and how to go beyond classical product coherent states. 

\begin{figure}
    \centering
    \includegraphics[width=1.0\columnwidth]{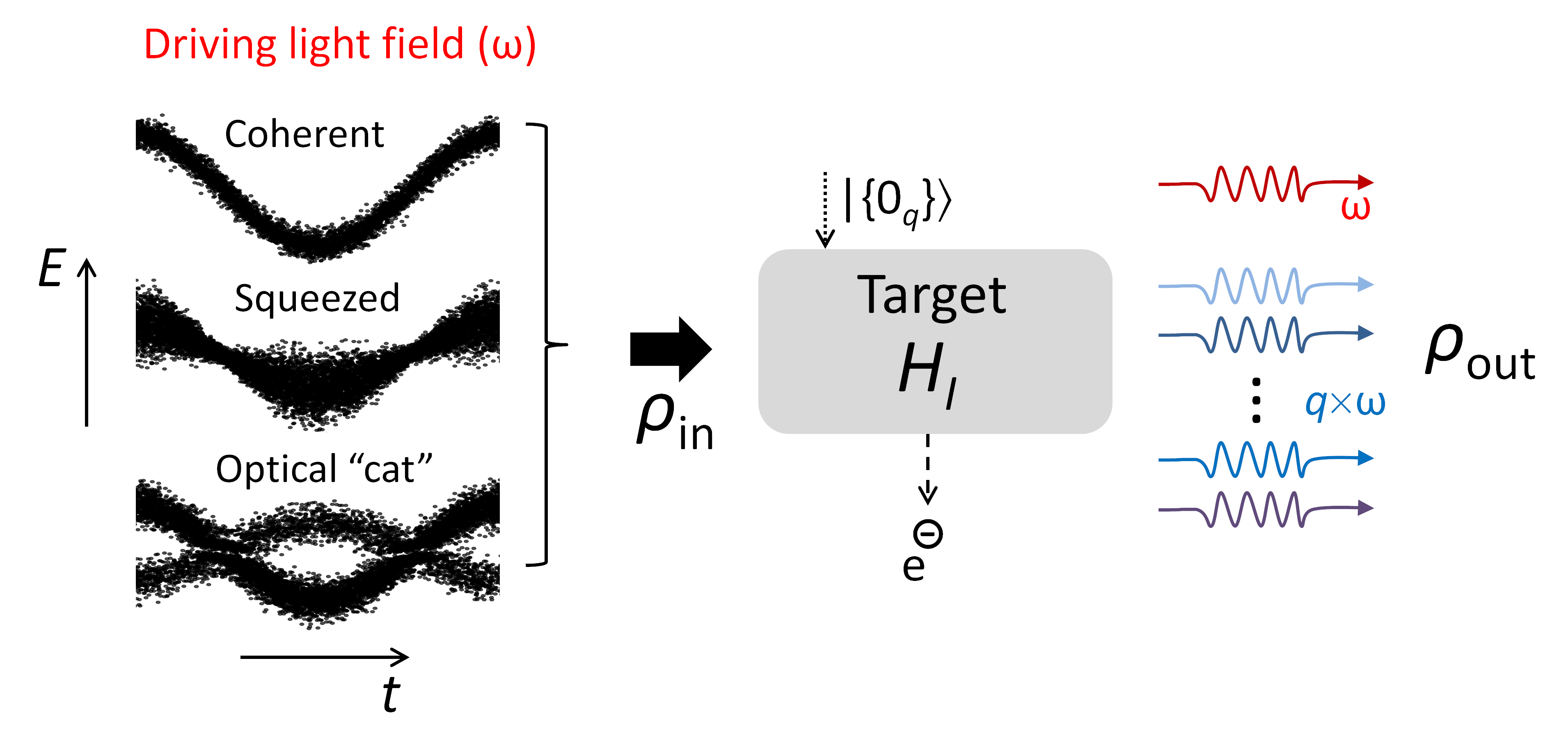}
    \caption{A schematic of the fully quantized description of intense laser--matter interaction. Here, as an example, we show the interaction of matter (which can be atoms, molecules or solids) with intense driving light fields of a coherent (laser), squeezed and an optical ``cat'' state of frequency $\omega$. In the driving fields we emphasize on the quantum noise of the different driving field distributions. $H_{I}$ is the interaction part of the Hamiltonian. The interaction products are electrons and photons including high harmonics of frequencies $\omega_{q}=q \omega$. $\rho_{in}=\ket{\Psi_{in}} \bra{\Psi_{in}}$ is the density operator describing the field state before the interaction, $\rho_{out}$ is the density operator describing the state of the system after the interaction and $\ket{ \{0_q\}}$ is the vacuum state of the harmonic modes $q$ before the interaction.}
    \label{fig:Fig_FullQuantum_RMP}
\end{figure}

Let us start by considering a single electron, described by the system Hamiltonian $H_S$, coupled to a quantized field within the dipole approximation~\cite{stammer2023quantum}
\begin{align}
\label{eq:hamiltonian}
    H = H_S + \sum_{q\ge 1} \hbar \omega_q a_q^\dagger a_q - \vb{d} \cdot \vb{E}_Q,
\end{align}
where $a_q^{(\dagger)}$ are the annihilation (creation) operators of the field in mode $q$. The initial state of the total system is given by $\ket{\Psi_i} = \ket{g} \otimes \ket{\Phi_i} $, with the electron initially in the ground state and the initial field state is given by $\ket{\Phi_i} = \ket{\alpha} \otimes \ket{\{ 0_q\}}$. 
Under typical experimental conditions the interaction of the electronic dipole moment $\vb{d}$ with the quantized field $\vb{E}_Q$ is well described by the dipole approximation, in which the electric field is treated without a spatial dependence, such that the light-matter coupling can be described via $H_I = - \vb{d} \cdot \vb{E}_Q$, see \cite{stammer2023quantum} for a detailed quantum electrodynamical derivation from the minimal coupling Hamiltonian. 

To arrive at a suitable form of the Hamiltonian we perform a set of unitary transformations, which allows to perform calculations in a frame in which further treatment is more convenient, and allows to introduce assumptions based on the strong field approximation. Note that the equations remain exact under unitary transformations. We first perform a unitary transformation with respect to the field Hamiltonian $H_F = \sum_q \hbar \omega_q a_q^\dagger a_q$, which adds a time dependence to the field operators $a_q \to a_q e^{- i \omega_q t}$, followed by a unitary displacement operation $D(\alpha) = \text{exp}[\alpha a_1^\dagger - \alpha^* a_1]$ on the driving field mode. The action of the displacement operation separates the classical from the quantum interaction via the relation $D(\alpha) a D^\dagger(\alpha) = a - \alpha$. 
What this transformation does, is to shift the initial driving laser field to the vacuum, and accounts for it with an additional interaction term in the Hamiltonian, in which the dipole moment is coupled to the classical electric field of the coherent driving laser. It is no accident that this is the exact same interaction Hamiltonian from the semi-classical picture $H_{I,sc} (t) = - \vb{d} \cdot \vb{E}_{cl}(t)$, where the classical field is properly defined as $\vb{E}_{cl}(t) = \Tr[\vb{E}_Q(t) \dyad{\alpha}]$. 
This allows to perform the final unitary rotation by which means we transform into the semi-classical frame of the electron. The system Hamiltonian $H_S$ in Eq.~\eqref{eq:hamiltonian}, together with the semi-classical interaction $H_{I,sc}(t)$, constitutes the well-known semi-classical Hamiltonian $H_{sc}(t) = H_S + H_{I,sc}(t)$ (see Sec.\ref{sec:semiclassical}). 
The interaction Hamiltonian is consequently given by 
\begin{align}
    H_I (t) = - \vb{d}(t) \cdot \vb{E}_Q(t),
\end{align}
where the $\vb{d}(t)$ is the time-dependent dipole moment operator of the electron and 
\begin{align}
    \vb{E}_Q(t) = -i g \sum_q \sqrt{q} \left(a_q^\dagger e^{i \omega_q t} - a_q e^{- i \omega_q t} \right),   
\end{align}
is the electric field operator with the light-matter coupling constant $g = \sqrt{\hbar\omega/(2V \epsilon_0)}$, and quantization volume $V$. The dipole moment operator is given in the semi-classical interaction picture
\begin{align}
    \vb{d}(t) & = U_{sc}^\dagger (t) \vb{d} U_{sc}(t), \\
    U_{sc}(t) & = \exp[- \frac{i}{\hbar} \int_0^t d\tau H_{sc}(\tau) ],
\end{align}
where $H_{sc}(t)$ is the semi-classical Hamiltonian given in Eq.~\eqref{eq:hamiltonian_semi}. 
The time evolution of the total state $\ket{\Psi (t)}$ is now given by the time-dependent Schrödinger equation 
\begin{align}
\label{eq:schrödinger}
    i \hbar \, \partial_t \ket{\Psi(t)} = H_I(t) \ket{\Psi(t)},
\end{align}
which can be solved formally for the total state $\ket{\Psi(t)} = U(t) \ket{\Psi_i}$. In general, due to the coupling between the dipole and the field, the field becomes entangled with the matter system, and, in addition, different field modes can also be entangled~\cite{stammer2022theory}. 
To make the matter-field correlations explicit, we introduce an identity on the electronic subspace $\mathds{1} = \dyad{g} + \int d \vb{v} \dyad{\vb{v}}$, such that the total state can be decomposed into terms corresponding to different electronic states 
\begin{align}\label{Eq:HHG:ATI}
    \ket{\Psi(t)} = K_{HHG} \ket{\Phi_i} \ket{g} + \int d\vb{v} K_{ATI}(\vb{v}) \ket{\Phi_i} \ket{\vb{v}}, 
\end{align}
where $K_{HHG} = \bra{g} U(t) \ket{g}$ and $K_{ATI}(\vb{v}) = \bra{\vb{v}} U(t) \ket{g}$.
These are the effective operators acting on the field state upon conditioning on different electronic states.~We can see that this describes an entangled state between light and matter decomposed in the basis of the electron. Ultimately, we want to find the solution of the total state $\ket{\Psi(t)}$, and particularly of the quantum optical field state. 
However, due to the complex dynamics, the Schrödinger equation in Eq.~\eqref{eq:schrödinger} can in general not be solved directly and one needs to find approximate solutions, making use of the characteristics of the intense driving field. 
For the process of HHG we can use the intuition from the semi-classical picture, in which the electron recombines with the parent ion, such that we consider the state of the field conditioned on the electron ground state $\ket{g}$. 
Doing so, this leaves us with a pure field state 
\begin{align}
    \bra{g} \ket{\Psi(t)} \equiv \ket{\Phi(t)} = \bra{g} U(t) \ket{g} \ket{\Phi_i},
\end{align}
and the problem of finding a solution reduces to finding an expression for the conditioned operation on the field state $\bra{g} U(t) \ket{g}$. 
The conditioned time-evolution operator acting on the field subspace is given by 
\begin{align}
    K_{HHG} = \bra{g} \mathcal{T} \exp[- \frac{i}{\hbar} \int_0^t d\tau H_I(\tau)] \ket{g},
\end{align}
which is a daunting expression to evaluate due to the commutator structure of the exponent, where $[H_I(t), H_I(t')] \in \mathcal{H}_S \otimes \mathcal{H}_F$ is an operator on the total Hilbert space. 
However, we can proceed to solve for the quantum state of the field by considering the Schrödinger equation for the conditioned field state
\begin{align}
    i \hbar \, \partial_t \ket{\Phi(t)} = \bra{g} H_I(t) \ket{\Psi(t)},
\end{align}
where we have used that $i \hbar\partial_t U(t) = H(t) U(t)$. To use the physical intuition about the underlying dynamics from the known semi-classical picture, and to apply pertinent approximations, we shall again use the identity on the electronic subspace
\begin{equation}\label{Eq:diff:atoms}
    \begin{aligned}
        i \hbar \, \partial_t \ket{\Phi(t)} = & \bra{g} H_I(t) \ket{g} \ket{\Phi(t)} \\
        & + \int d\vb{v} \bra{g} H_I(t) \ket{\vb{v}} \ket{\Phi(t,\vb{v})},
    \end{aligned}
\end{equation}
where $\ket{\Phi(t,\vb{v})} \equiv \bra{\vb{v}} \ket{\Psi(t)}$ is the field state conditioned on the continuum state of the electron. 

This expression is still exact and does therefore not provide simplifications, but writing the operator in differential form allows to perform approximations based on the well justified strong field assumptions~\cite{amini2019symphony}. First, we assume that the ground state of the atom is hardly depleted such that the second term in Eq.~\eqref{Eq:diff:atoms}, involving the population of continuum states, is very small compared to the first term $\norm{\ket{\Phi(t)}} \gg \norm{\ket{\Phi(t,\vb{v})}}$. We therefore only consider the homogeneous solution of the differential equation, given by 
\begin{align}
\label{eq:HHG_operation}
    K_{HHG} =  \mathcal{T} \exp[- \frac{i}{\hbar} \int_0^t d\tau \bra{g} H_I(\tau) \ket{g} ], 
\end{align}
where we can write the average of the interaction Hamiltonian over the atom ground state as
\begin{align}
    \mathrm{H}_I(t) = \bra{g} H_I(t) \ket{g} = - \bra{g} \vb{d}(t) \ket{g} \cdot \vb{E}_Q(t).
\end{align}

We find that the conventional dipole moment expectation value, pervasive in the semi-classical description, is now coupled to the field operator. Coupling the dipole expectation value $\bra{g} \vb{d}(t) \ket{g} = \expval{\vb{d}(t)}$, instead of the dipole operator, provided a substantial simplification of the daunting commutator structure. Now, the commutator $[ \mathrm{H}_I(t), \mathrm{H}_I(t') ] \in \mathds{C}$, is a c-number, which allows to find a closed form solution to Eq.~\eqref{eq:HHG_operation}. 
We emphasize that this is effectively a mean-field approach to the dipole moment operator, and essentially neglects all dipole moment correlations~\cite{sundaram1990high}.
Using that the interaction is now linear in the field operators, the solution is given by a multimode displacement operator 
\begin{align}
    K_{HHG} = \prod_q D[\chi_q(t)] \equiv \mathbf{D}[\mathbf{\bar \chi}],
\end{align}
where the coherent state amplitudes are given by the Fourier transform of the time-dependent dipole moment expectation value 
\begin{align}
\label{eq:coherent_amplitude}
    \chi_q (t) = g \sqrt{q} \int_{t_0}^t dt' \expval{\vb{d}(t')} e^{- i \omega_q t'}.
\end{align}

\begin{figure}
    \centering
    \includegraphics[width=0.8\columnwidth]{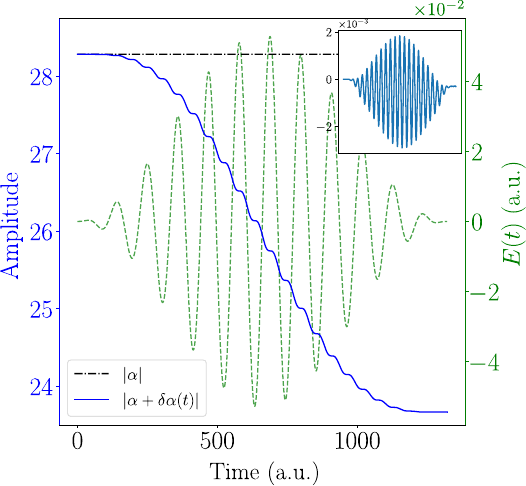}
    \caption{Depletion of the driving laser amplitude $\abs{\alpha + \delta \alpha (t)}$ from the initial amplitude $\alpha$ due to the strong field interaction and photon upconversion of the driving field into harmonic photons. The pulse of the driving laser field is shown (green, dashed) and the depletion (blue, solid) follows the ionization events of the electron. The inset shows how the amplitude's phase evolves over time. The figure has been reproduced from \cite{rivera2022strong}.}
    \label{fig:depletion}
\end{figure}

Consequently, the final field state is given by a product of Gaussian coherent states 
\begin{align}
\label{eq:state_coherent_product}
    \ket{\Phi(t)} = \ket{\alpha + \delta \alpha} \otimes \ket{\chi_2} \otimes ... \otimes \ket{\chi_{q_c}},
\end{align}
where we have transformed back into the original laboratory frame and defined $\chi_1 \equiv \delta \alpha$. 
This solution relies on the aforementioned picture that coherent input states map to coherent output states, which holds under the assumption of negligible dipole moment correlations, justified in the regime of small ground-state  depletion~\cite{stammer2024entanglement}. 
The emission of coherent radiation in the harmonics can be understood in terms of a classical charge current coupled to the quantum field, which leads to the generation of coherent states~\cite{scully1997quantum}. The classical charge current is given by the expectation value of the dipole moment, which encodes all the non-linearity of the strong field driven electron, and captures all the essential effects of the semi-classical approach to HHG. 
The shift in the fundamental driving mode $\delta \alpha$ accounts for the depletion of the driving laser due to the interaction with the HHG medium, and the transfer of energy from the pump to the harmonics~\cite{lewenstein2021generation, rivera2022strong}. The change in amplitude of the driving laser field, $\abs{\alpha + \delta \alpha(t)}$, is visualized in Fig.~\ref{fig:depletion}, which shows a decrease in the pump field amplitude following the cycles of the driving field.
It is also this solution which provides the closest resemblance to the established semi-classical picture of HHG, where the spectrum is obtained from the Fourier components of the dipole moment, just like in Eq~\eqref{eq:coherent_amplitude}. It is the same quantity considered here, and the product coherent state in Eq.~\eqref{eq:state_coherent_product} is essentially a classical result. Nevertheless, despite being in total agreement with semi-classical picture the quantum state of the field modes could be derived, and we have learned about the depletion of the pump field, which is usually assumed to be unaffected by the interaction.

The result of the quantum optical state of the harmonics in Eq.~\eqref{eq:state_coherent_product}, which is given by a product of coherent states, has found wide applicability in the description of quantum optical HHG. In particular, this result is used when going beyond coherent driving fields (see Sec.~\ref{sec:beyond_coherent}), in which the arbitrary driving field is decomposed in the coherent state basis and utilize the result that coherent input states are mapped to coherent output states. 
Here, we shall emphasize that there has been additional work on the quantum optical description of HHG in terms of single photon states in each harmonic mode~\cite{Gorlach2020}. These results show that including more internal bound electronic states beyond just the ground state leads to incoherent HHG spectra, lacking the well-defined structure predicted by semi-classical theory. In the subsequent sections we introduce the underlying origin and discuss the physical implications on the structure of the quantum optical state. We further note that there exist alternative results, following an Ehrenfest approach, which discuss the quantum optical signatures in HHG~\cite{de2024quantum}.

\subsection{Dipole correlations and second order interaction}
\label{sec:correlations}

We have seen that considering a mean field approach, by neglecting dipole moment correlations, allowed to find an approximate solution of the quantum state of the field in HHG. This solution was essentially classical in the sense that it was a product state over all modes, and each mode was in a coherent state. In the derivation of this result we have assumed negligible dipole moment correlations, which is justified in scenarios where only the ground state of the electron contributed to the dynamics. In the previous section, this was in the regime of vanishing ground state depletion, i.e. negligible excited state population. We shall now discuss the interesting consequences, and deviations from the classical mean-field result, if dipole correlations of the form $\expval{d(t) d(t')}$ are taken into account~\cite{stammer2024entanglement}. 
Therefore, instead of neglecting the second term in Eq.~\eqref{Eq:diff:atoms}, we explicitly include the contribution from ionization processes by considering
\begin{align}
    i \hbar \, \partial_t \ket{\Phi(t,\vb{v})} = &   \bra{\vb{v}} H_I(t) \ket{g} \ket{\Phi(t)} \\
    & + \int \vb{v}' \bra{\vb{v}} H_I(t) \ket{\vb{v}'} \ket{\Phi(t, \vb{v'})}. \nonumber
\end{align}
If we now neglect continuum-continuum transitions in the second term, we can approximate the solution by formal integration
\begin{align}
    \ket{\Phi(t,\vb{v})} = - \frac{i}{\hbar} \int_0^t dt' \bra{\vb{v}} H_I(t') \ket{g} \ket{\Phi(t^\prime)},
\end{align}
and substituting this into the second term in \eqref{Eq:diff:atoms} yields 
\begin{align}
    & i \hbar \, \partial_t \ket{\Phi(t)} = \bra{g} H_I(t) \ket{g} \ket{\Phi(t)} \\
    & - \frac{i}{\hbar} \int_0^t dt' \int d\vb{v} \bra{g} H_I(t) \ket{\vb{v}} \bra{\vb{v}} H_I(t') \ket{g} \ket{\Phi(t')}. \nonumber
\end{align}

While we have now effectively eliminated the ionization processes in the solution, but explicitly keep transitions to the continuum in the matrix elements $\bra{g}H_I(t) \ket{\vb{v}}$, we still have a non-trivial integration kernel such that further approximations are necessary. We therefore perform a Markov type approximation in which we replace $\ket{\Phi(t')} \to \ket{\Phi(t)}$, and further consider only terms in second order of the interaction. This allows to write the operation on the quantum state as~\cite{stammer2024entanglement, lange2024hierarchy}
\begin{align}
\label{eq:state_squeezed}
    \ket{\Phi(t)} = \mathbf{D}[\mathbf{\bar \chi}] \exp[-\frac{1}{2} \bra{g} Q^2(t) \ket{g}] \ket{\{ 0_q \}},
\end{align}
where we have defined 
\begin{align}
    Q(t) \equiv \frac{1}{\sqrt{\hbar}} \int_{t_0}^t dt' \left[ H_I(t') - \bra{g} H_I(t^\prime) \ket{g} \right].
\end{align}

We can directly see by inspection that the additional term is in second order of the interaction $H_I(t)$, which will ultimately lead to quadratic terms in the operation acting on the field state. Such quadratic terms are responsible for squeezing along the field modes~\cite{schleich2001phase}, and we will now see which crucial electronic property determines the squeezing of the field. 
For the quadratic term we can write
\begin{align}
\label{eq:q-term}
    \expval{Q^2(t)} = & \int_{t_0}^t \int_{t_0}^t dt' dt'' \, \vb{E}_Q(t') \vb{E}_Q(t'') \\
    & \times \left[ \expval{\vb{d}(t') \vb{d}(t'')} - \expval{\vb{d}(t')} \expval{\vb{d}(t'')} \right]. \nonumber
\end{align}

\begin{figure}
    \centering    
    \includegraphics[width=1\columnwidth]{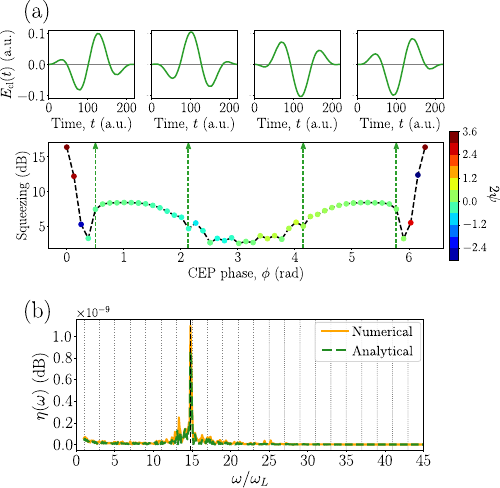}
    \caption{(a) Squeezing parameter of the fundamental field mode after HHG interaction as a function of the relative phase (CEP) between the carrier wave and the envelope (shown in the upper panels for specific CEP values indicated by arrows). The color coding indicates the squeezing angle. (b) Squeezing parameter of the harmonic spectrum for a driven Mott insulator. At the resonance frequency of the exciton, the squeezing shows dominant features in the non-classical response. The figures have been reproduced from (a)~\cite{stammer2024entanglement}, and (b)~\cite{lange_excitonic_2025}.}
    \label{fig:squeezing}
\end{figure}

\begin{figure}
    \centering
    \includegraphics[width=1\columnwidth]{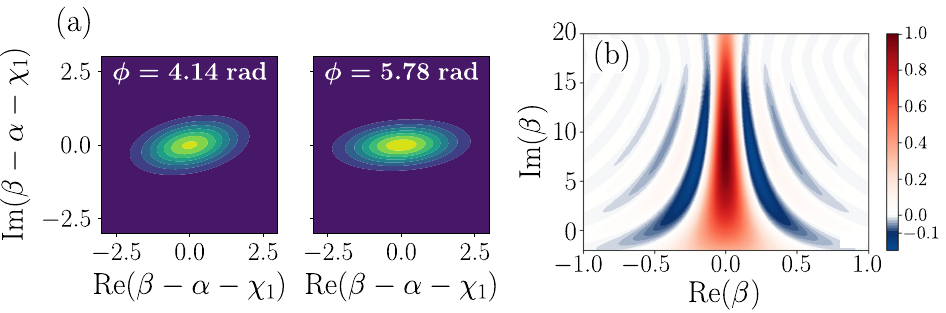}
    \caption{Wigner function of squeezed states after HHG. (a) Squeezing along different quadrature directions for the fundamental laser field after the HHG interaction. The two phases indicate the CEP (compare with Fig.~\ref{fig:squeezing}~(a)). In (b) the Wigner function of the 3rd harmonic showing squeezing like signatures, as well as Wigner negativities, for HHG in a cavity with population of AC-shifted resonances. The figures have been reproduced from (a)~\cite{stammer2024entanglement}, and (b)~\cite{Misha2025resonant}. }
    \label{fig:squeezing_wigner}
\end{figure}

We can see that non-trivial contributions from the second order correction only appear if dipole-moment correlations $\expval{\vb{d}(t') \vb{d}(t'')}$ are non-negligible~\cite{stammer2024entanglement}. 
Since the field operator $\vb{E}_Q$ appears twice in this contribution, it causes an effective squeezing operation acting on the field modes. In Fig.~\ref{fig:squeezing}~(a) the squeezing of the fundamental mode due to the dipole moment correlations is shown, together with its ability to control the squeezing phase by varying the phase of the carrier wave to the envelope of the driving field pulse.
It was also shown that such correlations in the electron current of a Mott Insulator can lead to enhanced squeezing signatures in the presence of exciton resonance energies~\cite{lange_excitonic_2025}. The enhanced squeezing at the exciton energy is shown in Fig.~\ref{fig:squeezing}~(b).
We further emphasize that the presence of squeezing signatures in the HHG field state is not restricted to continuum state populations as outlined above. The effect is universal once the quantum electrodynamical interaction is going beyond the mean-field approach of the dipole. This was seen in \cite{Gorlach2020, rivera2024squeezed}, where instead of continuum states, the population of bound electronic states was taken into account. Or likewise in \cite{Misha2025resonant}, when AC-Stark shifted resonances appear during the interaction of the field with the matter system. 
These characteristics can be seen in the Wigner functions of the light field, such as the squeezing in the fundamental mode after the HHG interaction~\cite{stammer2024entanglement}, or the non-classical signatures after resonant excitation~\cite{Misha2025resonant}, as shown in Fig.~\ref{fig:squeezing_wigner}~(a) and (b), respectively. 
These results on the generation of squeezed harmonics are the immanent quantum features in HHG when going beyond the mean field approach of the dipole. 

Besides the generation of squeezing in the harmonic field modes, the second order interaction due to dipole moment correlations can cause the generation of entangled photon pairs. While HHG is dominated by the dipole moment expectation value $\expval{d(t)}$, the second order contribution from the dipole correlations $\expval{d(t) d(t')}$ can generate photon pairs of energy $\hbar \omega$ and $\hbar \omega'$, with the constraint from energy conservation in the in the conversion process~\cite{sloan2023entangling}.

\subsection{Coherence in quantum HHG }
\label{sec:coherence}

In the previous two sections we have discussed the approach to quantum optical HHG via the Schrödinger picture, in which the quantum state of the field was propagated in time under specific assumptions about the electron dynamics. We shall now discuss a complementary approach, given by the Heisenberg picture~\cite{stammer2025theory}. Here, we propagate the operators instead of the state. 
Solving for the time-dependent field operators has the advantage that quantum optical properties, such as optical coherence or two-time correlations functions, can be computed directly.

The Heisenberg equation of motion (EOM) for the field operators is given by~\cite{diestler2008harmonic, stammer2025theory, sundaram1990high}
\begin{align}
\label{eq:heisenberg_EOM}
    \dv{t} a_q(t) & = \frac{i}{\hbar} \left[ H(t), a_q(t) \right] \\
    & = - i \omega_q a_q(t) + g \sqrt{q} \sum_{i=1}^N d_i(t),
\end{align}
where $H(t)$ is the Hamiltonian of interest, such as the one given in Eq.~\eqref{eq:hamiltonian}. 
The generic solution to the EOM is accordingly given by 
\begin{align}
\label{eq:heienberg_solution}
    a_q(t) = & \, a_q e^{- i \omega_q t} \\
    & + g \sqrt{q} \int_0^t dt' e^{- i \omega_q (t-t')} \sum_{i=1}^N d_i(t'), \nonumber
\end{align}
where the first term is the free evolution of the field, while the second term corresponds to the source term due to the presence of $N$ charges.~While the solution to the Heisenberg EOM is in principle exact, the strong field dynamics allows to separate the quantum from the classical dynamics (akin to the discussion in Sec.~\ref{Sec:Quant:state}), such that for the time-dependent dipole moment in the Heisenberg picture only the dynamics under the intense classical field is considered~\cite{diestler2008harmonic, stammer2025theory}. This allows to explicitly solve for the field quantities considered below. Equipped with the time-dependent field operators in the Heisenberg picture, one can now compute the first and second order field correlation functions~\cite{mandel1995optical}, which are related to the spectrum of the field and the photon statistics, respectively. 
For the first order field correlation function
\begin{align}
    G^{(1)}(t,t+\tau) = \expval{a_q^\dagger (t) a_q(t+\tau)},    
\end{align}
we can separate the correlations into a coherent and an incoherent contribution by using
\begin{align}
\label{eq:dipole_decomposition}
    \expval{d(t_1) d(t_2)} = \expval{d(t_1)} \expval{d(t_2)} + \expval{\Delta d(t_1) \Delta d(t_2)},
\end{align}
where $\Delta d(t) = \expval{d(t)} - d(t)$ are the dipole fluctuations around its mean. 
Since the dipole moment operator can in general fluctuate, the HHG spectrum should in principle not only be considered from the mean value of the dipole $\expval{d(t)}$, as done in semi-classical theory or other approaches to quantum optical HHG~\cite{lewenstein2021generation}. But instead, the stochastic component of the dipole operator should be considered. 
The Wiener-Khintchine theorem (WKT) relates the power spectral density $S(\omega)$ of a fluctuating process to the first order auto-correlation function~\cite{wiener1930generalized, khintchine1934korrelationstheorie}. For a stationary random process these two are a Fourier transform pair~\cite{mandel1995optical, carmichael2013statistical}, and thus $S(\omega)$ is given by 
\begin{align}
    S(\omega) = \frac{1}{\pi} \operatorname{Re} \left[ \int_0^\infty d\tau \lim_{t \to \infty} G^{(1)}(t,t+\tau) e^{i \omega \tau} \right].
\end{align}

\begin{figure}
    \centering
    \includegraphics[width=1\columnwidth]{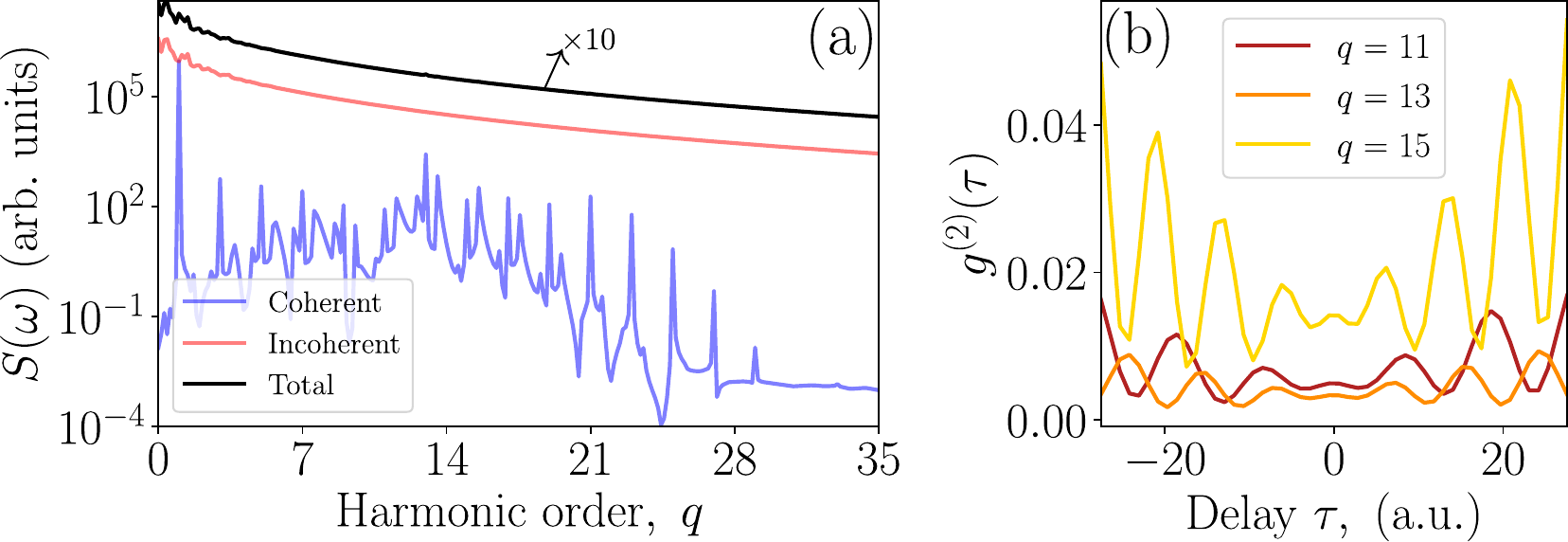}
    \caption{(a) Total spectrum of HHG obtained from the Wiener-Khintchine theorem $S(\omega) = S_{coh}(\omega) + S_{inc}(\omega)$, i.e.~as a sum of the coherent and incoherent contributions (black). While the coherent contribution shows the conventional peak and plateau structure of HHG (blue), the incoherent contribution has no structure (red). Interestingly, the incoherent contribution is orders of magnitude larger than the coherent emission, however, due to the different scaling with the number of emitters ($N$ vs. $N^2$), the coherent contribution dominates in the experimental response of many uncorrelated emitters. (b) Second order correlation function $g^{(2)}(\tau)$ of different harmonics in the single atom scenario ($q \in \{ 11,13,15 \}$). The second order correlation function indicates clear signatures of photon anti-bunching in the emission characteristics of the harmonics ($g^{(2)}(0) < 1$), making it the only genuine non-classical photon statistics in HHG. The figures have been reproduced from \cite{stammer2025theory}.}
    \label{fig:coherence}
\end{figure}

With the decomposition of the dipole moment correlation in the first order field correlation function into a mean value and its fluctuations, see Eq.~\eqref{eq:dipole_decomposition}, the power spectrum decomposes into a coherent and an incoherent contribution $S(\omega) = S_{coh}(\omega) + S_{inc}(\omega)$. 
While the coherent contribution is the well known Fourier transform of the dipole expectation value
\begin{align}
    S_{coh}(\omega) = \hbar^2 g^4 q^2 \abs{\expval{d(\omega_q)}}^2 \delta (\omega - \omega_q),
\end{align}
the incoherent contribution is obtained from transition matrix elements between ground and continuum states~\cite{stammer2025theory}
\begin{align}
    S_{inc}(\omega) =  \hbar^2 g^4 q^2 \delta(\omega - \omega_q) \int dv d_{gv}(\omega_q) d_{vg}^*(\omega_q), 
\end{align}
where $d_{gv}(\omega_q) = \int dt \bra{g} d(t) \ket{v} e^{- i \omega_q t}$. 
Interestingly, the single emitter contribution of the incoherent part can be larger than the non-linear contributions from the oscillations of the dipole moment expectation value as indicated in Fig.~\ref{fig:coherence}~(a). However, in the many emitter scenario, it was shown that the coherent contribution scales as $N^2$, while the incoherent contribution only scales as $N$ \cite{stammer2025theory, Gorlach2020, sundaram1990high}. Thus, observing these contributions becomes challenging in conventional HHG experiments, where a large number of atoms interact with the driving field.
Given the first order correlation function $g^{(1)}(t,t+\tau)$, it is now straightforward to show that the HHG process possess first order optical coherence in the long time limit~\cite{stammer2025theory}
\begin{align}
    g^{(1)}(\tau) \equiv  \lim_{t \to \infty } \frac{G^{(1)}(t,t+\tau)}{\abs{G^{(1)}(t,t) G^{(1)}(t+\tau, t+\tau)}^{1/2}} = 1.
\end{align}

However, given the time-dependent field operator allows a straightforward calculation of the normalized second order field correlations function
\begin{align}
    g^{(2)}(\tau) = \frac{\expval{:I(t) I(t+\tau):}}{\expval{I(t)} \expval{I(t+\tau)}},
\end{align}
where $::$ indicates normal ordering of the field operators, and $I(t) \propto a^\dagger_q(t)a_q(t)$ is the intensity operator of the $q$-th harmonic field mode. The second order field correlation function can therefore be seen as an intensity correlation measurement. 
Considering classical light fields, in which the intensity does not follow any commutation relations, the following inequalities help to differentiate between classical and non-classical light properties in the intensity correlation measurement. For classical fields it holds that $g_{cl}^{(2)}(0) \ge g_{cl}^{(2)}(\tau)$ and $g_{cl}^{(2)}(0) \ge 1$. The property of the normalized second order correlation function also allows to classify the photon statistics of the light field for vanishing time delay $\tau =0$, and for $g^{(2)}(0) > 1$ and $g^{(2)}(0) < 1$ is referred to photon bunching and photon anti-bunching, respectively~\cite{mandel1995optical}. For $g^{(2)}(0) = 1$ there are no correlations and the photons are detected at random, following the typical Poisson statistics of coherent radiation. Among these, only the anti-bunching signature of photons is a witness of non-classical light fields~\cite{Kimble1977}, while the other photon statistics have classical counterparts.
Note that photon anti-bunching and sub-Poissonian photon statistics are closely related, but that one does not imply the other since sub-Poissonian statistics can be accompanied by photon bunching~\cite{zou1990photon}. 
Calculating the $g^{(2)}(\tau)$ function for the HHG process has predicted anti-bunching signatures in the emission statistics of the harmonics~\cite{stammer2025theory}, and is shown in Fig.~\ref{fig:coherence}~(b). These are one of the first clear predictions of non-classical photon statistics in HHG, in contrast to classical bunching properties in the harmonics observed in~\cite{lemieux2024photon}.

With the development of the Heisenberg EOM for the field operators, the extension to other scenarios is straightforward.~For instance, including an environment consisting of independent bosons with Hamiltonian $H_E = \int dk \, \hbar \omega_k r_k^\dagger r_k$, where $r_k^{(\dagger)}$ are the environmental annihilation (creation) operators, the interaction Hamiltonian with the environment reads~\cite{stammer2025quantum}
\begin{align}
    H_{I,E} = \int d k \sum_q \left( g_k a_q^\dagger r_k + g_k^* a_q r_k^\dagger \right),
\end{align}
with mode-environment couplings $g_k$.~The additional Hamiltonian $H_{I,E}$ will consequently add additional terms in the Heisenberg EOM of Eq.~\eqref{eq:heisenberg_EOM}, which in the Markovian case\footnote{Assuming that the environment is Markovian, i.e. the quantum noise from the environment and associated damping does not lead to memory effects, is obtained when considering an unstructured environment (white noise) where the coupling is independent of the mode $g_k \approx g_0$. } leads to a damping of rate $\kappa \propto g_0^2$, i.e.~proportional to the coupling of the environment~\cite{stammer2025quantum}. The EOM, in the context of open and stochastic dynamics often called quantum Langevin equation, is now given by 
\begin{align}
    \dv{t} a_q(t) = & \left[ - i \omega_q - \kappa \right] a_q(t) + g \sqrt{q} \, d(t) \\
    & - i g_0 \int dk \, r_k(0) e^{- i \omega_k t}, \nonumber
\end{align}
where the system-environment coupling is independent of the mode ($g_k \approx g_0$), due to the Markov approximation.
Computing the spectrum in the presence of the continuum of environment modes from the WKT, we find for the coherent part of the $q$-th harmonic mode
\begin{align}
    S_{coh}(\omega) = \hbar^2 g^4 q^2 \sum_{N} \frac{\abs{\expval{d(\omega_N)}}^2}{(\omega_q - \omega_N ) + \kappa^2} \, \delta(\omega - \omega_N),
\end{align}
where $\omega_N$ are the Fourier frequencies of the dipole moment expectation value $\expval{d(\omega_N)}$. This expression shows that the presence of an environment leads to a spectral profile of peaks with natural linewidth and a Lorentzian shape, showing the isomorphism between a non-linear antenna and quantum optical HHG in the mean field approximation. It furthermore allows to compute the emitted power by the incoherent contribution from the emitter fluctuations $P_{max} \propto g^2/g_0^2 $, which is bounded by the ratio of the coupling constants between light-matter and light-environment~\cite{stammer2025quantum}.

\subsection{Entanglement in HHG}\label{Sec:Entanglement}

Given the product coherent state description for the high harmonic field modes under the mean-field approach derived in Sec.~\ref{Sec:Quant:state} \cite{lewenstein2021generation, rivera2022strong}, it momentarily questions the presence of entanglement between the field modes. However, in Sec.~\ref{sec:correlations} we have seen that this result was suitable for solutions under specific assumptions, and when including dipole correlations already goes beyond the coherent state description.
It was shown that going beyond the mean field approach, and including fluctuations of the dipole moment results in quadratic interaction terms. These additional contributions not only lead to squeezing in the field, but also result in a correlation of all field modes~\cite{stammer2024entanglement}. This can be seen when explicitly writing down in Eq.~\eqref{eq:q-term} the field operators 
\begin{align}
    \expval{Q^2} = & - g^2 \sum_{q,p} \sqrt{qp} \left[ \mathcal{G}_{qp} a_q^\dagger a_p^\dagger  - \mathcal{K}_{qp} a_q^\dagger a_p  + \operatorname{h.c.} \right],
\end{align}
which is a generic bilinear hermitian Hamiltonian for the bosonic field operators~\cite{stammer2024entanglement}. The coefficients $\mathcal{G}_{qp}$ and $\mathcal{K}_{qp}$ form the matrix elements including the dipole correlations. 

\begin{figure}
    \centering
    \includegraphics[width=1.0\columnwidth]{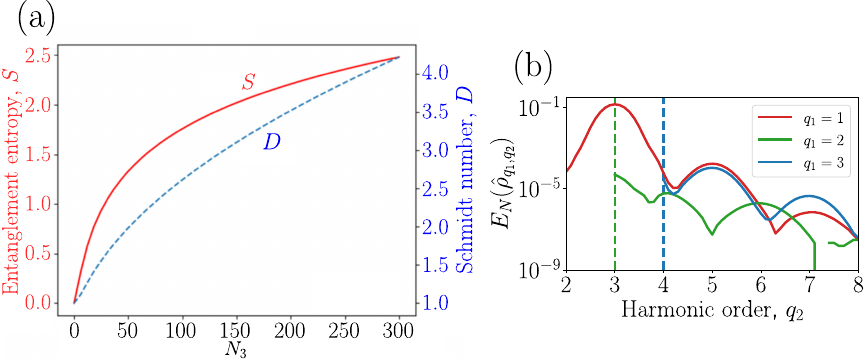}
    \caption{(a) Entanglement entropy $S$ and Schmidt number $D$ for the entanglement of the third and fifth harmonic in resonant HHG, for increasing average photon number. The increase in both entanglement measures indicates that the entanglement increases for increasing quantum state size. (b) Logarithmic negativity for different harmonic pairs in HHG, starting from an initial excited state. Panels adapted from (a) \cite{Misha2025resonant} and (b) \cite{rivera2024squeezed}.}
    \label{fig:entanglement}
\end{figure}

Such higher order interaction, and therefore field correlations, can be induced by specifically shaping the atomic emitter. 
For instance, in Eq.~\eqref{eq:state_squeezed} it was demonstrated that including electronic continuum states in the dynamics leads to the presence of the second order corrections. It is therefore natural to expect that traces of bound-state populations are imprinted on the field properties in the form of entanglement. In fact, it was predicted that initial electronic coherence~\cite{rivera2024squeezed} or resonant bound state population~\cite{Misha2025resonant} leads to field entanglement. 
In particular, Fig.~\ref{fig:entanglement}~(a) shows the entanglement entropy $S$ and the Schmidt number $D$ for the entangled state between the third and fifth harmonic~\cite{Misha2025resonant}, both growing for increasing average photon number, indicating that entanglement increases with the size of the state (in terms of photon number). Further, Fig.~\ref{fig:entanglement}~(b) shows the logarithmic negativity for different pairs of harmonic orders~\cite{rivera2024squeezed}. 
Alternatively, distinguishing between the short and long trajectories contributing to HHG reveals that separating the individual trajectories allows for field correlations~\cite{rivera2024role}. 
This can be understood that distinguishing between different contributions from the electronic states (apart from the ground state), will each contribute to different quantum optical states of the harmonics. But since all these contributions appear coherently, the final field state will be entangled. 

In addition to these theoretical advances in understanding the origin of field entanglement in HHG, there are first experimental results performed in semiconductor driven HHG~\cite{theidel2024evidence, theidel2024observation}, which report a violation of a Cauchy-Schwarz inequality (CSI) in the measured intensity correlation functions. It is known that classical light fields obey $g^{(2)}_{ii} g^{(2)}_{jj} \le [g^{(2)}_{ij}]^2$ for the normalized second order intensity correlation function $g^{(2)}$. A violation of this inequality is a direct witness of field correlations, and can be quantified in the parameter
\begin{align}
    R \equiv \frac{[g^{(2)}_{ij}]^2}{g^{(2)}_{ii} g^{(2)}_{jj}},
\end{align}
where entanglement between the modes $i$ and $j$ is observed when $R > 1$. A violation of the CSI for the harmonic orders $q \in \{3,5\}$ was measured in~\cite{theidel2024evidence}. However, note that these results do not unambiguously reflect the properties of the quantum state of the light field itself, as the non-stationary nature of the pulsed radiation influences the measurement statistics in a non-trivial way~\cite{van2025errors}.

\subsection{Quantum HHG beyond simple systems }
\label{sec:complex}

While in atomic systems the HHG process mainly involves the interplay between a non-degenerate ground and continuum states, in more complex systems, such as molecules or solids, the coupling between multiple electronic states plays an important role. In such cases, the identity inserted in Eq.~\eqref{Eq:diff:atoms} does not solely reduce to the ground and continuum states, but may include additional bound states. We thus consider the generic resolution of the identity in some eigenbasis of the system under consideration $\mathds{1} = \sum_{i} \dyad{\psi_i}$, leading to a more general system of coupled differential equations
\begin{equation}\label{Eq:Complex:Systems}
	\begin{aligned}
	i\hbar \, \partial_t \ket{\Phi_i(t)}
		&= \langle \psi_i \vert H_I(t) \vert \psi_i \rangle
			\ket{\Phi_i(t)}
			\\& \quad 
			+ \sum_{j \neq i}
				\langle \psi_i\vert H_I(t) \vert \psi_j \rangle
					\ket{\Phi_j(t)}
			\ \forall i,
	\end{aligned}
\end{equation}
with $\ket{\Phi_i(t)} \equiv \bra{\psi_i}\ket{\Psi(t)}$, and reduces to Eq.~\eqref{Eq:diff:atoms} when restricted to the strong-field approximation in atomic HHG, i.e. only considering the atomic ground and continuum states.

In these systems, however, the ground-continuum-ground pathway may not be the only, or even the dominant, channel contribution to the harmonic emission. For example, in molecules, recombination through excited electronic states can become significant, depending on the molecular structure and symmetry~\cite{lein_mechanisms_2005,bian_multichannel_2010,bian_phase_2011}. In solids, where electronic wavefunctions are naturally delocalized across bands, recombination can occur at different points within a band with nonzero probability~\cite{parks_wannier_2020}. In such cases, the basis set $\{\ket{\psi_i}\}$ is naturally described either in terms of Bloch states~\cite{vampa_theoretical_2014} or, when a more atomic-like interpretation is desirable, Wannier states~\cite{osika_wannier-bloch_2017,parks_wannier_2020}.

\begin{figure}
    \centering
    \includegraphics[width=1\columnwidth]{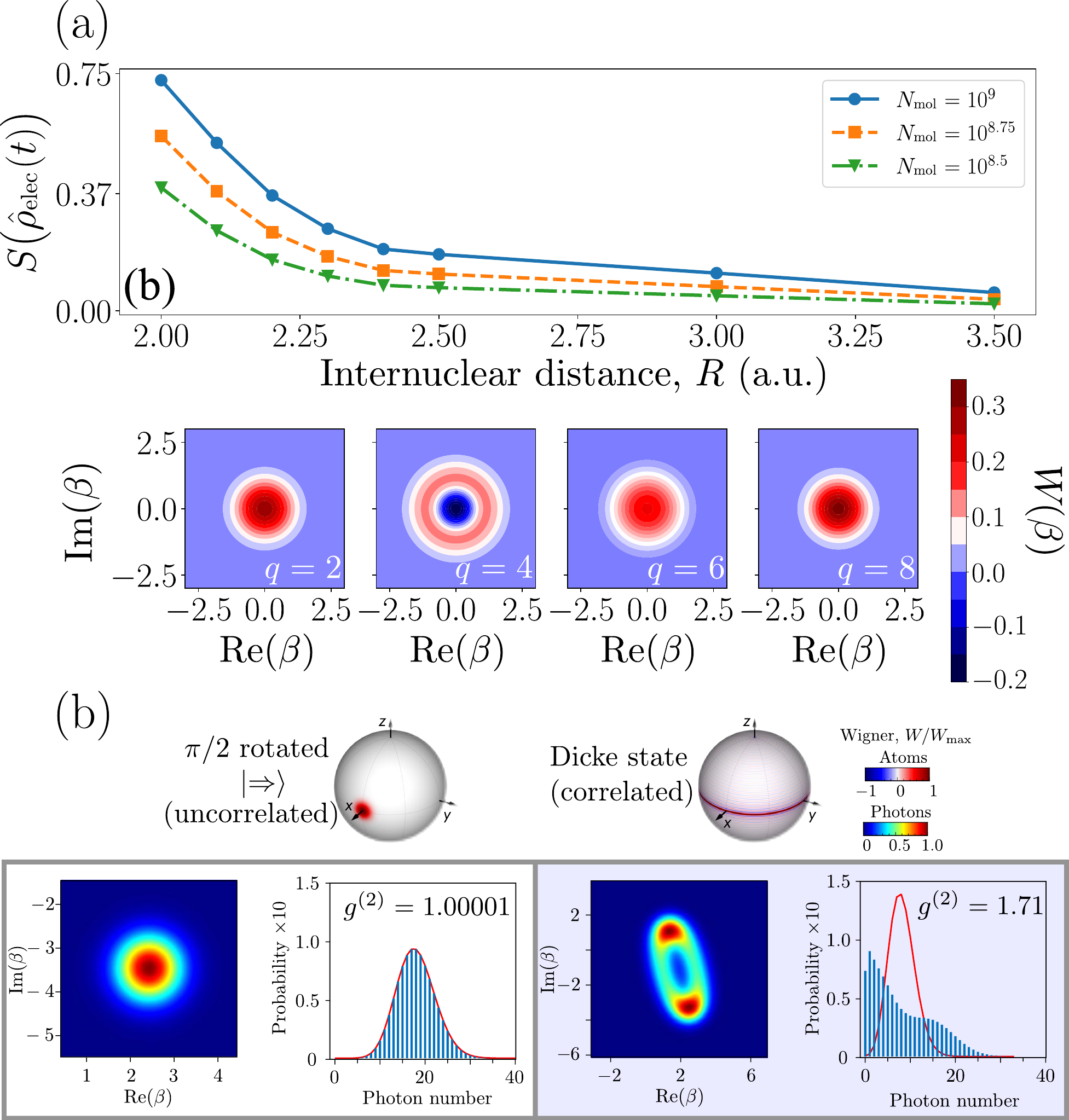}
    \caption{(a) Light-matter entanglement after HHG in H$^{+}_2$ molecular ions as a function of the internuclear distance, shown for different numbers of particles in the interaction region. The lower panel displays the Wigner functions of several harmonic modes when postselecting the electron in the antibonding state.~(b) Wigner function and photon statistics of the 21st harmonic order after HHG in a many-body atomic system, for initially uncorrelated (left panel) and quantum correlated (right panel) atomic ensembles. Nevertheless, all results in panel (b) remain classical. Panels adapted from (a) \cite{rivera2024molecules} and (b) \cite{pizzi2023light}.} 
    \label{fig:many:body}
\end{figure}

Recombination into a state different from the initial one can give rise to non-classical light-matter correlations, which can be used to tailor the final quantum optical state.~In the simple case of diatomic molecules, such as the H$_2^+$ molecular ion, where the bonding (ground) and anti-bonding (first excited) state correspond to symmetric and antisymmetric superpositions of the atomic ground states of the respective atomic nuclei, it has been shown that the interplay between electronic localization and delocalization across the two-centers can induce light-matter entanglement features on the post-interaction state~\cite{rivera2024molecules} (see Fig.~\ref{fig:many:body}~(a)). In particular, postselecting on events in which the electron recombines into the bonding state, with the electron delocalized between the two nuclei, yields coherent state radiation, analogous to the atomic case. In contrast, recombination into the antibonding state, with the electron localized around one of the two atomic centers, can leave the harmonics in a non-Gaussian state (see Fig.~\ref{fig:many:body}~(a)), since such transitions are mediated by single-photon excitations. This behavior, however, strongly depends on the internuclear distance: as the bond length increases, the probability of ionization at one center followed by recombination at the other decreases, thereby suppressing cross terms of the form $\langle \psi_i \vert H_I(t)\vert \psi_j(t)\rangle$, with $i\neq j$, that favor inversion of population between ground and excited states.

The influence of non-parametric recombinations and electronic localization in many-body systems on the non-classical features of emitted HHG radiation can also be illustrated by studying transitions between highly localized electronic phases and more delocalized ones. This is in particular the case of the Mott-insulator to metallic phase transition in half-filled Fermi-Hubbard models~\cite{lange2024electron,lange2024hierarchy}. In the half-filled Fermi-Hubbard model the number of electrons equals the number of lattice sites:~electrons can hop between neighboring sites with amplitude $t$, but if two electrons of opposite spin occupy the same site, they experience an on-site repulsion $U$.~At large ratios $U/t$, the system enters a Mott-insulating phase, where double occupancy is strongly suppressed, and each site is typically occupied by a single electron. Excitations of this configuration can be generated by a strong laser field, producing doublons (doubly occupied sites) and holons (empty sites) quasiparticles, which can contribute to harmonic generation~\cite{silva_high-harmonic_2018}.~In this regime, $\langle\psi_i\vert H_I(t)\vert\psi_j \rangle \neq 0$ for $i\neq j$, where $\ket{\psi_i}$ represents the electronic state at site $i$.~This leads to non-classical features in the emitted harmonics, observed as small squeezing effects~\cite{lange2024electron,lange2024hierarchy}, which become more pronounced when excitons, i.e., bound doublon-holon pairs, are formed in the system~\cite{lange_excitonic_2025} (see Fig.~\ref{fig:squeezing}~(b)).~By contrast, in the metallic phase, $\langle \psi_i \vert H_I(t) \vert \psi_j \rangle = j_{ii}(t) \delta_{ij}$, decoupling the dynamics in Eq.~\eqref{Eq:Complex:Systems}, result in harmonic modes given by coherent states.

In semiconductor materials, the electrons contributing to HHG are initially delocalized in the valence band. Unlike in metallic phases, they can undergo the standard three-step process characteristic of HHG: excitation to the conduction band, acceleration in the field, and recombination back in the valence band~\cite{vampa_theoretical_2014}.~In this case, two main mechanisms contribute to the generation of high-order harmonics: intraband dynamics, corresponding to electron oscillations within a given band, and interband dynamics, corresponding to transitions between different bands, with the latter dominating the higher-order harmonics response.~Under low depletion conditions of the valence band, the final state of the harmonics is analogous to that of the atomic case, see Eq.~\eqref{eq:state_coherent_product}, with the displacement given by $\chi_q(t) = \chi^{(\text{intra})}_q(t)+\chi^{(\text{inter})}_q(t)$, which includes contributions from both intra- and interband electron dynamics~\cite{rivera2024nonclassical}. When focusing solely on intraband dynamics, while still accounting for the joint electron-light backaction during the interaction, signatures of squeezing in the emitted radiation have also been reported~\cite{gonoskov2024nonclassical}.~Along this line, recent experimental studies in semiconductors have demonstrated the presence of multimode displaced squeezed states in the low-order harmonic regime~\cite{theidel2024evidence,theidel2024observation}.

However, not only the intrinsic features of the matter system can significantly affect the quantum optical properties of the harmonic emission, but also the initial state in which the system is prepared.~For example, preparing atomic systems in a many-body entangled state, such as a Dicke-state\footnote{The Dicke-state is the equal superposition of an $N$ atom man-body state, where $m$ atoms are in the excited state $\ket{e_i}$, and the remaining $N-m$ atoms are in the ground state $\ket{\psi} \propto \sum_{\pi} \ket{e_1,...,e_m,g_{m+1},...,g_{N}}_{\pi}$, with the sum running over all different permutations.}, it has been shown to substantially modify the harmonic emission, leading to super-Poissonian photon statistics with non-trivial, though positive Wigner functions for single harmonic modes (see Fig.~\ref{fig:many:body}~(b)), and to classical correlations shared among them~\cite{pizzi2023light}. Although these states are non-Gaussian, the results in~\cite{pizzi2023light} do not possess any quantum signatures due to the Hudson's theorem~\cite{hudson1974wigner}.
In contrast, genuine non-classical features in the form of multimode squeezed light can emerge when the HHG process is initiated from a first-excited state~\cite{rivera2024squeezed}, as discussed in Sec.~\ref{Sec:Entanglement}.

An alternative many-body system for generating high-order harmonics, involving ultracold Bose gases, has recently been studied~\cite{stammer2025high}. Based on the quantum field theory of atoms interacting with photons~\cite{lewenstein1994quantum, you1995quantum, you1996quantum}, it was shown that the generated quantum optical state of the harmonics after the interaction with a Bose-Einstein condensate show genuine non-classical signatures, by virtue of entanglement and squeezing across all field modes. Furthermore, above the critical temperature $T_c$ for Bose-Einstein condensation, the emitted quantum state is a mixture of Gaussian states. 
While high harmonic emission from many-body systems can give rise to non-Gaussian quantum states, as in the case of Dicke-states~\cite{pizzi2023light} or Bose gases, these signatures are not quantum, and only the Bose-Einstein condensate emits true quantum light~\cite{stammer2025high}.

\subsection{Conditioning on HHG and generation of optical Schrödinger cat states}
\label{sec:conditioning}

Given the diverse theoretical progress of the field when integrating quantum optics and strong-field processes, the first experimental evidence for the generation of non-classical light states using the HHG process was presented in \cite{lewenstein2021generation}. The reconstructed quantum state resembled the one of an optical Schrödinger cat state, which has been further refined in purity with enhanced Wigner negativity~\cite{rivera2022strong}, and even been extended to high photon numbers sufficient to drive non-linear processes~\cite{lamprou2025nonlinear}.
This was achieved using conditioning and post-selection techniques applied to the interaction products of the HHG process in atoms~\cite{lewenstein2021generation, rivera2022strong, stammer2023quantum, tsatrafyllis2017high, stammer2022high, stammer2022theory, Stammer_PRA_2025, Stammer_EnergyConservation2024, gonoskov2016quantum}, and theoretically extended to molecules, solids or the process of Thomson scattering by free electrons~\cite{rivera2024molecules, rivera2024solids, Andrianov_ThomosonScat_PRA_2025}.

\begin{figure}
    \centering
    \includegraphics[width=0.99\columnwidth]{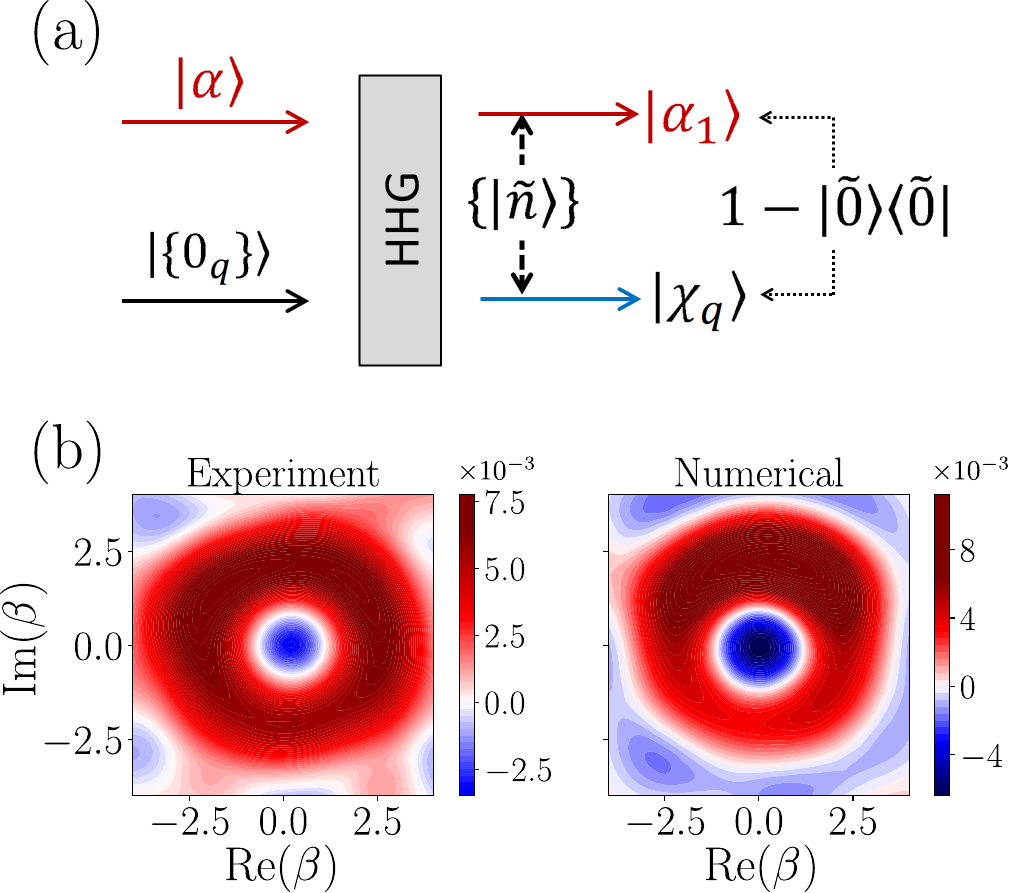}
    \caption{(a) Conditioning scheme for the generation of optical cat states. $\ket{\alpha }$ and $\ket{\alpha + \delta \alpha }$ are the IR states before and after the interaction, respectively, with $|\alpha + \delta \alpha |<|\alpha|$ due to energy conservation. The harmonics are in the vacuum $\ket{0_{q}}$ and coherent states $\ket{\chi_{q}}$ before and after the interaction, respectively. The wave-packet modes to capture the correlations between the field amplitudes are depicted as $\ket{{\Tilde{n}}}$, with $\mathds{1}-\dyad{\Tilde{0}}$ being the post-selection operator applied for the creation of the optical cat states~\cite{stammer2022high, stammer2022theory}. (b) Comparison between the experimental data of conditioning in HHG with a numerical sampling approach of the shot-to-shot measurement statistics and post-selection on energy conserving events. The fidelity between these two Wigner functions is $F=0.93$, and both exhibiting pronounced Wigner negativities indicating the non-classical signatures of the state. Panel (b) was adapted from \cite{Stammer_PRA_2025}. }
    \label{fig:Conditioning1_scheme}
\end{figure}

The generation of the optical cat states conceptually relies on post-selection techniques on the energy-conserving events in the shot-to-shot measurement statistics of the harmonic radiation and the driving field~\cite{Stammer_PRA_2025}. This can be approximately described as a projection on the outgoing state from the medium~\cite{stammer2022high, stammer2022theory}.
Since the interaction with the material system preserves energy, photon exchange across different frequency modes produces correlations among the field modes and in the measured photon statistics. To capture these correlations theoretically, one can formally introduce a set of wave-packet modes $\{\ket{\tilde{n}}\}$, with $\tilde{n}=\tilde{0}$ representing the absence of HHG excitations. These quasi-excitations of modes naturally connect the physical harmonic modes because the excitations originate from the same energy conserving electron dynamics. Then, the conditioning procedure can be described by the set of positive operator-valued measures (POVM) $\{\mathds{1}-\dyad{\Tilde{0}},\dyad{\Tilde{0}}\}$, which represent whether or not the HHG process and the corresponding infrared (IR) laser depletion have occurred~\cite{stammer2022theory}. We refer to $\mathds{1}-\dyad{\Tilde{0}}$ as the conditioning on HHG (see Fig.~\ref{fig:Conditioning1_scheme}~(a) for a schematic representation). Applying this operator to the outgoing HHG state $\ket{\alpha + \delta \alpha} \otimes_{q}\ket{\chi_{q}}$, see Eq.~\eqref{eq:state_coherent_product}, we create a massively entangled state between all field modes~\cite{stammer2022high}
\begin{equation}\label{Eq:HHG:cond1}
    \ket{\Phi_{\text{IR,HHG}}}
        = \ket{\alpha + \delta \alpha }
            \bigotimes^{N_{\text{c}}}_{q=2}
                \ket{\chi_q}
            -\xi_1 \ket{\alpha} \bigotimes^{N_{\text{c}}}_{q=2}
                \xi_q\ket{0_q},
\end{equation}
where $\xi_{1}=\langle \alpha | \alpha + \delta \alpha \rangle$ and $\xi_{q}= \langle 0_q| \chi_q\rangle$.
The generation of the optical \textit{cat} state in the IR spectral region can then be obtained by projecting the state $\ket{\Phi_{\text{IR,HHG}}}$ onto the harmonic state. In this case, one obtains  
\begin{equation}\label{Eq:HHG:cond}
    \ket{\Phi_{\text{IR}}} = \ket{\alpha + \delta \alpha }
            - \bra{\alpha} \ket{\alpha + \delta \alpha} e^{- \Omega} \ket{\alpha},
\end{equation}
where $\Omega = \sum_{q \ge 2} \abs{\chi_q}^2$. This expression corresponds to a generalized optical cat state. 

\begin{figure}[b]
    \centering
    \includegraphics[width=0.99\linewidth]{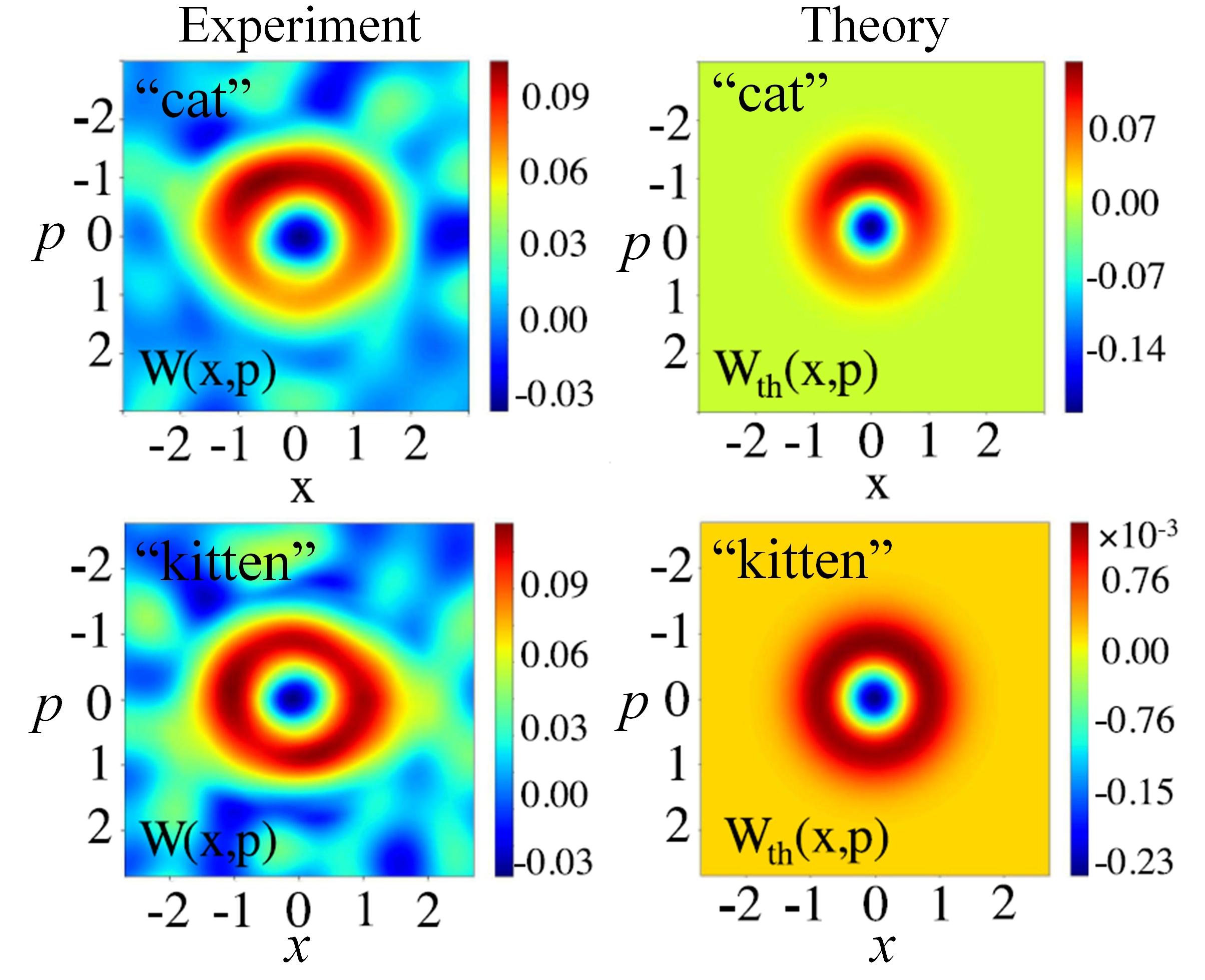}
    \caption{The left panel shows the Wigner function $W(x,p)$ of the experimentally measured IR optical cat (large $|\delta\alpha|$) and ``kitten'' (small $|\delta\alpha|$) states generated by conditioning on the HHG process induced by intense laser--atom interactions. The right panel shows theoretically calculated Wigner functions of the corresponding states. The figure has been reproduced from \cite{rivera2022strong}. }
    \label{fig:kitacat}
\end{figure}

The experimental results of the corresponding state tomography measurements, given by the reconstructed Wigner function, together with the theoretical calculations are shown in Fig.~\ref{fig:kitacat}. The quantum features of the state can be controlled by adjusting the atomic density in the interaction region, which affects the coherent state shift $\delta\alpha$ \cite{rivera2022strong}. The upper panel of Fig.~\ref{fig:kitacat} corresponds to the generation of an optical cat state (large $|\delta\alpha|$), while the lower panel shows an optical ``kitten'' state obtained by reducing the gas density in the HHG region, i.e. small $|\delta\alpha|$. 
Although experimental limitations in the conditioning approach currently restrict the fidelity of the measured IR optical cat states to approximately 70\%, theoretical simulations under experimentally similar conditions~\cite{Stammer_PRA_2025, Stammer_EnergyConservation2024} have demonstrated that the post-selection scheme can generate optical cat states with fidelities above 90\% [Fig.~\ref{fig:Conditioning1_scheme}~(b)].~These theoretical post-selection results have been obtained via numerical sampling~\cite{Stammer_PRA_2025}, where the shot-to-shot measurement statistics of the IR-XUV modes is sampled from the distribution given by the respective quantum state and post-selected on energy-conserving events was performed. 

Extending this approach to generate non-classical optical states in extreme wavelength regimes, the HHG process enables entanglement across optical field modes, ranging from the far--IR to XUV regime~\cite{stammer2022high, stammer2022theory}. Specifically, interchanging the role of the fundamental with the harmonics in the field state Eq.~\eqref{Eq:HHG:cond1}, and subsequently measuring the harmonic modes $q'\neq q$, we can obtain a superposition of a coherent state with the vacuum in the XUV spectral regime~\cite{stammer2022high}. 

A direct consequence of the post-selection approach is that it can be used for introducing optical cat states obtained from HHG to non-linear optics. This has been demonstrated in \cite{lamprou2025nonlinear} by developing a decoherence-free approach, which leads to the generation of intense fs-duration IR optical cat states. These states have been utilized in non-linear optics by driving the second harmonic generation process in an $\beta$--barium borate (BBO) crystal. It has been shown that the non-classical features of the intense IR cat sates are imprinted in the second order autocorrelation traces of the second harmonic, and on the quantum features of the light state of the generated second harmonic.
This approach can be used for creating non-classical light states in various spectral regions via nonlinear up-conversion processes, and provides a way of characterizing high photon number optical cat states in regimes where conventional state tomography methods reach their limitations.

\subsection{Quantum optical HHG beyond coherent driving fields}
\label{sec:beyond_coherent}

In the previous sections, we have primarily focused on scenarios where the initial quantum optical state was described by classical states. This choice is well-justified, as the attosecond community has made significant advances over the past three decades in generating high-intensity, classical coherent laser fields, which have become an elementary source in strong-field experiments. Nevertheless, recent developments in quantum optical light engineering has brought the increasing attention for generating high-intensity non-classical light sources, particularly bright squeezed vacuum (BSV) states with mean photon numbers as high as $10^{13}$~\cite{spasibko2017multiphoton,manceau2019indefinite-mean}. These states have demonstrated substantial enhancements in multi-photon optical harmonic generation and supercontinuum generation~\cite{spasibko2017multiphoton,manceau2019indefinite-mean}, owing to their heavy-tailored photon number distributions, as to their ultrafast intensity fluctuations for broadband BSV.

Thus, the possibility of generating non-classical states of light in the high-photon number regime naturally raises the question of how such states might affects the HHG process and its properties. An elegant approach to address this question, building upon the formalism presented in~\cite{lewenstein2021generation}, see Sec.~\ref{Sec:Quant:state}, was proposed in~\cite{gorlach2023high}, where the initial state of the driving field is expressed using the generalized positive $P$-representation~\cite{drummond1980generalised}
\begin{equation}
	\rho_L(t_0) 
		= \int \dd^2 \alpha \int \dd^2 \beta \,
				\dfrac{P(\alpha,\beta^*)}{\braket{\beta^*}{\alpha}}
				\dyad{\alpha}{\beta^*},
\end{equation}
such that the initial joint light-matter state reads $\rho(t_0) = \dyad{\text{g}}\otimes \rho_L(t_0) \otimes \dyad{\{0_q\}}$.~Here, $P(\alpha,\beta^*)$ is the phase-space distribution that encapsulates all the quantum properties of the driving field.~Unlike other standard phase-space representations, such as the Glauber-Sudarshan $P$-representation~\cite{schleich2001phase}, which expresses quantum optical states as $\rho = \int \dd^2\alpha P(\alpha) \dyad{\alpha}$, the generalized positive $P$-representation uses a double phase-space integral. This key distinction avoids the singularities that can arise in $P(\alpha)$ for certain non-classical states, such as squeezed states, and instead allows to define smooth, well-behaved and real-valued $P(\alpha,\beta^*)$ functions~\cite{drummond2016quantum}. Furthermore, the representation in terms of the coherent state basis allows one to express, in the low-depletion regime, the joint light-matter state at a given time $t$ as
\begin{equation}\label{Eq:state:gen:P}
	\begin{aligned}
	\rho(t)
		= \int \dd^2 \alpha
			\int \dd^2 \beta
				\dfrac{P(\alpha,\beta^*)}{\braket{\beta^*}{\alpha}}
				&\dyad{\psi_{\alpha}(t)}{\psi_{\beta^*}(t)}
				\\& \otimes
					\dyad{\Phi_{\alpha}(t)}{\Phi_{\beta^*}(t)},
	\end{aligned}
\end{equation}
where $\ket{\psi_{\alpha}(t)}$ represents the quantum state of the electron evolved under the semiclassical Hamiltonian with electric field $E_{cl}^{(\alpha)}(t) \propto \alpha \sin(\omega t)$~\cite{stammer2024limitations}, and $\ket{\Phi_\alpha(t)}$ is given by the product of coherent states from Eq.~\eqref{eq:state_coherent_product}. Note that, in contrast to the treatment in Sec.~\ref{Sec:Quant:state}, the coherent state amplitudes $\alpha$ and $\beta$ are here integration variables, not fixed parameters. This arises due to the representation of the arbitrary initial driving field in the coherent state decomposition. 

\begin{figure}
    \centering
    \includegraphics[width=1\columnwidth]{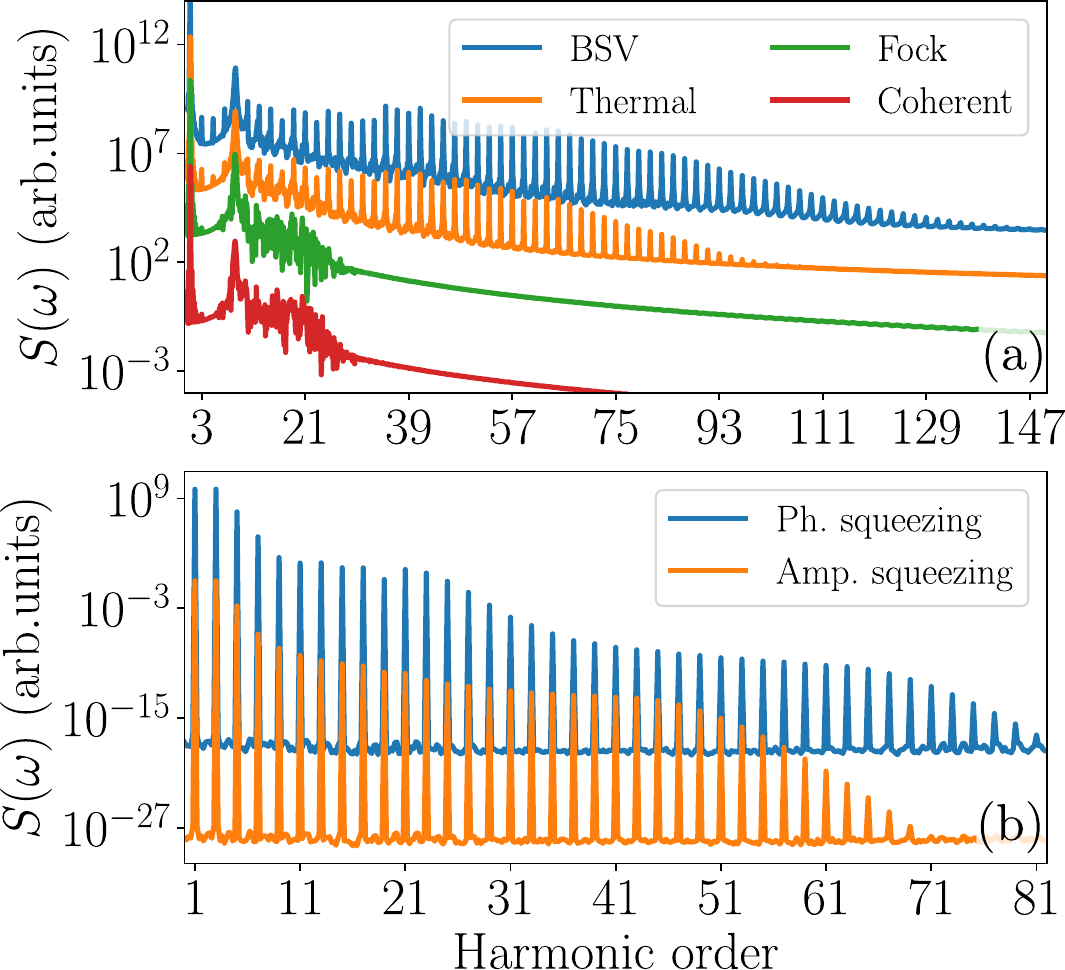}
    \caption{HHG spectrum computed for:~(a) various initial sources, where squeezed and thermal states yield higher HHG cutoffs than the others, while still being a purely classical signature; and (b) a circularly polarized driving field, where one polarization component exhibits squeezing features either in phase (blue) or in amplitude (orange). Panels adapted from~\cite{gorlach2023high} and \cite{rivera2025structured}, respectively.}
    \label{Fig:HHG:Spec:QLight}
\end{figure}

By means of Eq.~\eqref{Eq:state:gen:P}, gaining insights into how the photon statistics of the initial driving field affect the HHG outcomes, becomes straightforward, particularly for the HHG spectrum, which is evaluated as $S(\omega) = \lim_{g \to 0} \pdv*{\mathcal{E}(t)}{\omega}$ with $\mathcal{E}(t) = \sum_q \tr[a^\dagger_q a_q \rho(t)]$. Here, the limit $g\to 0$ corresponds to taking the quantization volume $V\to \infty$, which in turn requires $\alpha \to \infty$ in order to keep the electric field amplitude finite $E_{cl}(t) \propto \alpha / V$. These two conditions, jointly referred to as the \emph{classical limit}, are motivated by the fact that we are modeling laser-matter interactions in free space, where the quantization volume is effectively very large. It was estimated in~\cite{gorlach2023high} that, for typical strong-field parameters, this regime already holds for $V \approx 10^{-27}$ m$^3$. In addition, this approximation was further validated in~\cite{wang2025highorder}, where a different approach was adopted: the initial state of the driver was expressed using a discrete, yet (over)complete~\cite{bargmann1971completeness,perelomov1971completeness}, set of coherent states, $\ket{\Phi(t_0)} = \sum_{m,n}c_{m,n} \ket{\alpha_{m,n}} \otimes \ket{\{0_q\}}$. Beyond this formalism, it is important to highlight the contributions of~\cite{boroumand2025quantum}, who provided a microscopic analysis of HHG, where squeezed light acts as a perturbative addition to a coherent state driver; of~\cite{wang2024high}, who evaluated how the presence of a squeezed environment on the harmonic modes affects the HHG emission; and of~\cite{imai2025electron}, who explored more generally how the electron dynamics are influenced when the driving field is prepared in an optical Schrödinger cat state.

Under these considerations, and in particular when using single-mode Gaussian states of light, the evaluation of the HHG spectrum reduces to
\begin{equation}
	S(\omega) \propto
		\int \dd^2\alpha 
			\Big[ \lim_{g\to 0} P(\alpha,\alpha)\Big]
				\abs{\chi_{q}(t,\alpha)}^2,
\end{equation}
where $P(\alpha,\alpha)$ is a Gaussian distribution, whose widths determine the amount of field fluctuations present in the driving field and their distribution across the quantum optical phase space. For instance, in the case of an initial coherent state of amplitude $\alpha_0$, the above limit yields $\lim_{g\to\infty} P(\alpha,\alpha) = \delta(\alpha-\alpha_0)$, consistent with what would be obtained from Eq.~\eqref{eq:state_coherent_product}. For states exhibiting enlarged field fluctuations along one or both optical quadratures, such as squeezed or thermal states, the limiting distribution does not reduce to a Dirac delta function and, as a result, the presence of these fluctuations becomes directly imprinted on the HHG spectrum. Nevertheless, and independent of the quantum nature of the driving field, these signatures remain classical, since any classical distribution with large intensity fluctuations would yield the same characteristics in the HHG spectrum. 

In particular, it was predicted that driving the process with BSV or thermal light results in extended cutoff frequencies when compared to coherent states of equal mean intensity~\cite{gorlach2023high} [Fig.~\ref{Fig:HHG:Spec:QLight}~(a)], effectively showing that the use of enhanced field fluctuations can non-trivially modify the HHG electron dynamics~\cite{even2024motion}, and further lead to extremely short attosecond pulses~\cite{wang2025attosecond}. This was explicitly demonstrated in the context of elliptically polarized fields~\cite{rivera2025structured}, where the addition of squeezing features along one of the polarization components of the field enables HHG processes, in what are otherwise considered classically forbidden configurations~\cite{budil1993influence,antoine1996theory}. In this case, the HHG spectrum becomes sensitive to the optical quadrature along which squeezing occurs [Fig.~\ref{Fig:HHG:Spec:QLight}~(b)], with the observed modifications intuitively understood by recognizing that enhanced field fluctuations effectively act as an additional force on the electron during its acceleration in the continuum~\cite{even2023photon}. This alters the kinetic energy gained by the electron and enables recombination events even under circular polarization configurations. Interestingly, while these results were predicted when using atoms, similar effects have been reported in single-band semiconductors driven by strong-field quantum light, where the HHG mechanisms stems from intraband transitions~\cite{gothelf2025high}. 

Experimental developments have, however, kept pace with theoretical advances, and significant efforts have been devoted to using bright squeezed states of light, at intensities around $\sim 10^{12}$ W/cm$^2$, either as primary drivers of HHG in semiconductors~\cite{rasputnyi2024high}, or as auxiliary fields assisting strong coherent fields in both solid-state~\cite{lemieux2024photon} and gaseous media~\cite{tzur2025measuring}. When BSV states are used as the main driving field in semiconductor materials~\cite{rasputnyi2024high}, it has been demonstrated the possibility of employing mean intensities exceeding the material's damage threshold by over an order of magnitude, thereby opening new opportunities for probing solid-state systems in regimes inaccessible to intense coherent driving light. Alternatively, when using squeezed light as perturbative fields with a frequency distinct from that of a strong-coherent driver~\cite{lemieux2024photon,tzur2025measuring, stammer2025weak}, breaking the symmetry of the HHG process, it allows for harmonic sidebands depicting super-Poissonian photon statistics. While such statistics do not constitute a definitive signature of non-classicality, they nonetheless demonstrate the potential of this configuration for selective quantum state engineering of specific harmonic orders~\cite{tzur2024generation, stammer2025weak}. 

While the progress in the investigation of HHG driven by squeezed light sources has revealed interesting insights into the underlying dynamics, results on phase-randomized classical driving fields has shown that the harmonic spectrum does not reveal the presence of quantum optical coherence by means of off-diagonal density matrix elements in the eigenbasis~\cite{stammer2024absence}. 

\subsection{Propagation and decoherence effects of quantum light}
\label{sec:propagation}

Non-classical states of light, particularly those with high photon numbers, are highly sensitive to decoherence effects, which naturally arise from photon losses. Such losses may result from linear processes such as scattering or single-photon absorption. Photon losses reduce the state’s fidelity and, if they are strong, they can even completely destroy the quantum nature of the state.
This is particularly important for intense quantum light sources, where in addition to the linear processes, the highly nonlinear response of the material plays a significant role.

Investigating this problem requires the development of fully quantized approaches that account for the propagation of intense quantum light in strongly nonlinear media. This issue has recently attracted attention and has been investigated in the context of intense BSV sources propagating through gaseous media~\cite{rivera2025propagation}. It was shown that photon losses and decoherence effects, induced by strong-field processes during propagation in the medium, impose significant limitations on the nonlinear observables and the propagation length for which the field maintains its quantumness.
It was further predicted that the classical cut-off extension of HHG, predicted for the ideal single atom case~\cite{gorlach2023high}, is dismissed when considering realistic scenarios including propagation and decoherence effects~\cite{rivera2025propagation}.

\section{Quantum optical ATI}
\label{sec:ATI_classical}

\subsection{Quantum field coupled to ATI electrons}
\label{sec:ATI_part1}

The process of above-threshold ionization (ATI) generally refers to the ionization of an electron via the absorption of more photons than required to overcome the ionization potential of the corresponding atomic or molecular system~\cite{milosevic_above-threshold_2006,lewenstein_principles_2008,agostini_chapter_2012}. This definition spans both the multiphoton~\cite{agostini_free-free_1979} and tunnelling~\cite{chin_observation_1983,hansch_resonant_1997} phenomena, and for this reason ATI is often regarded as the first step towards HHG~\cite{fu2001interrelation}. Here, we refer to ATI as comprising all ionization events that lead to at least one electron remaining in the continuum. This broad picture encompasses a variety of processes~\cite{amini2019symphony}, from events where the ionized electron does not return to the parent ion (direct ATI)~\cite{Keldysh1965,chin_observation_1983}, passing through others where the electron is driven back and undergoes an elastic collision that modifies its momentum (high-order ATI)~\cite{lewenstein_rings_1995,paulus_above-threshold_2000,salieres_feynmans_2001}, to events where the re-colliding electron transfers enough energy to ionize a second electron (non-sequential double ionization)~\cite{lhuillier_multiply_1983,corkum_plasma_1993}.

Thus, although ATI can in general involve multi-electron dynamics, in this subsection we restrict our discussion to the single-electron cases, and more specifically to direct ATI (dATI) events. This leaves out other single-electron processes such as high-order ATI (HATI) which, despite the existence of proposed quantum-optical descriptions~\cite{stammer2023quantum}, have not yet had their broader implications analyzed in detail, thereby leaving an open avenue for further investigations. In contrast, a focused study of dATI provides a clear foundation for understanding how ATI processes affect the properties of the quantum optical field state~\cite{rivera2022light}, and how they can potentially give rise to strong radiation emission with probabilities orders of magnitude larger than those encountered in HHG~\cite{milovsevic2023quantum}.

Building on the quantum optical theory of HHG presented in Sec.~\ref{Sec:QHHG:revisited}, and noting that in ATI the relevant final electron state corresponds to electrons detected in the continuum with momentum $\ket{\vb{v}}$, we obtain from Eq.~\eqref{Eq:HHG:ATI} that the quantum optical state conditioned on finding an electron with momentum $\ket{\vb{v}}$ reads
\begin{equation}
	\braket{\vb{v}}{\Psi(t)}
		= \ket{\Phi(\vb{v},t)}
		= \bra{\vb{v}} U(t) \ket{g} \ket{\Phi_i},
\end{equation}
which evolves according to the effective time-dependent Schrödinger equation
\begin{equation}\label{Eq:diff:eq:ATI}
	\begin{aligned}
	i\hbar \, \partial_t \ket{\Phi(\vb{v},t)}
		&= \bra{\vb{v}}H_I(t)\ket{g} \ket{\Phi(t)}
			\\& \quad
			+ \int \dd \vb{v}' \bra{\vb{v}} H_I(t) \ket{\vb{v}'} \ket{\Phi(\vb{v}',t)}.
	\end{aligned}
\end{equation}
The second term in this expression describes electronic transitions within the continuum between different momentum states, and can be written as~\cite{rivera2022light}
\begin{equation}
	\begin{aligned}
	 &\int \dd \vb{v}' \bra{\vb{v}} H_I(t) \ket{\vb{v}'} \ket{\Phi(\vb{v}',t)}
	 	\\&\hspace{1.5cm}
	 	= \Delta \vb{r}(\vb{v},t)\cdot \vb{E}_Q(t) \ket{\Phi(\vb{v},t)}
	 	 \\&\hspace{1.5cm}\quad
	 	  + \int \!\!\dd \vb{v}' \, \vb{G}(\vb{v},\vb{v}',t)\cdot \vb{E}_Q(t) \ket{\Phi(\vb{v}',t)},
	 \end{aligned}
\end{equation}
where $\Delta \vb{r}(\vb{v},t)$ denotes the electronic displacement in the continuum between the ionization time $t_1$ up to a final time $t$~\cite{rivera2022light,stammer2023quantum}, given by
\begin{equation}
	\Delta \vb{r}(\vb{v},t,t_1) 
		= \dfrac{e}{m} \int^{t}_{t_1} \dd \tau \,
				\big[
					m \vb{v} 
					- e \vb{A}(t)
					+ e \vb{A}(\tau)					
				\big],
\end{equation}
and couples to the electric field operator as an oscillating charge current~\cite{ScullyBookCh2}. In contrast, $\vb{G}(\vb{v}',\vb{v})$ accounts for transitions between two momentum states mediated by the atomic potential, thereby characterizing re-scattering events~\cite{lewenstein_rings_1995,amini2019symphony}.

While each atomic and molecular system constitutes a world of its own, in general dATI events dominate the behavior of electrons emitted with kinetic energy smaller than or on the order of $2U_p$~\cite{milosevic_above-threshold_2006,amini2019symphony}, where $U_p$ denotes the ponderomotive potential. In this regime, the contribution of $\vb{G}(\vb{v},\vb{v'})$ responsible for HATI can be treated perturbatively. Accordingly, the zeroth-order contribution to Eq.~\eqref{Eq:diff:eq:ATI} reads
\begin{align}
	i\hbar \, \partial_t \relaxket{\Phi_{\text{dATI}}(\vb{v},t)}
		&= \bra{\vb{v}}H_I(t)\ket{g} \ket{\Phi(t)}
		\\& \quad
		+\Delta \vb{r}(\vb{v},t)\cdot \vb{E}_Q(t) \relaxket{\Phi_{\text{dATI}}(\vb{v},t)}, \nonumber
\end{align}
whose solution, assuming that the electron is initially in the ground state, can be expressed as
\begin{equation}\label{Eq:QO:dATI}
	\relaxket{\Phi_{\text{dATI}}(\vb{v},t)}
		= - \dfrac{i}{\hbar}
				\int^{t}_{t_0}
					\dd t'
						\vb{D}[\bar{\delta}(\vb{v},t,t')]
						\bra{\vb{v}}H_I\ket{g} \ket{\Phi(t')}.
\end{equation}

From this expression we see that, in addition to the classical electron dynamics encoded in $\bra{\vb{v}}H_I\ket{g}$, the charge current induced by the electron's propagation in the continuum induces a displacement, $\vb{D}[\bar{\delta}(\vb{v},t,t')]$, of the field modes. This displacement is given by
\begin{equation}
	\delta_q(\vb{v},t,t')
		= g\sqrt{q} 
			\int^t_{t_0} \dd \tau \,
				\boldsymbol{\epsilon}_q \cdot \Delta \vb{r}(\vb{v},\tau)
					e^{-i\omega_q \tau},
\end{equation}
which is the Fourier transform of the charge current generated by the propagating electron. Interestingly, the amount of optical displacement varies with the ionization conditions, determined both by the ionization time $t'$ and by the kinetic momentum of the released electron $\vb{v}$. Thus, when focusing on the direct ATI component of the joint light-matter state, we find
\begin{equation}
	\relaxket{\Psi_{\text{dATI}}(t)}
		\propto \int\! \dd \vb{v}\int^{t}_{t_0}\!\dd t'
			\vb{D}[\bar{\delta}(\vb{v},t,t')]
				\bra{\vb{v}}H_I(t')\ket{g} \ket{\Phi(t')},
\end{equation}
which in general corresponds to an entangled state between light and matter. Moreover, projection onto specific values of momentum, leading to Eq.~\eqref{Eq:QO:dATI}, yields quantum optical states that, to a good approximation~\cite{rivera2022light}, can be described as superpositions of coherent states with amplitudes dependent on the electron's ionization conditions.

\begin{figure}
    \centering
    \includegraphics[width=1\columnwidth]{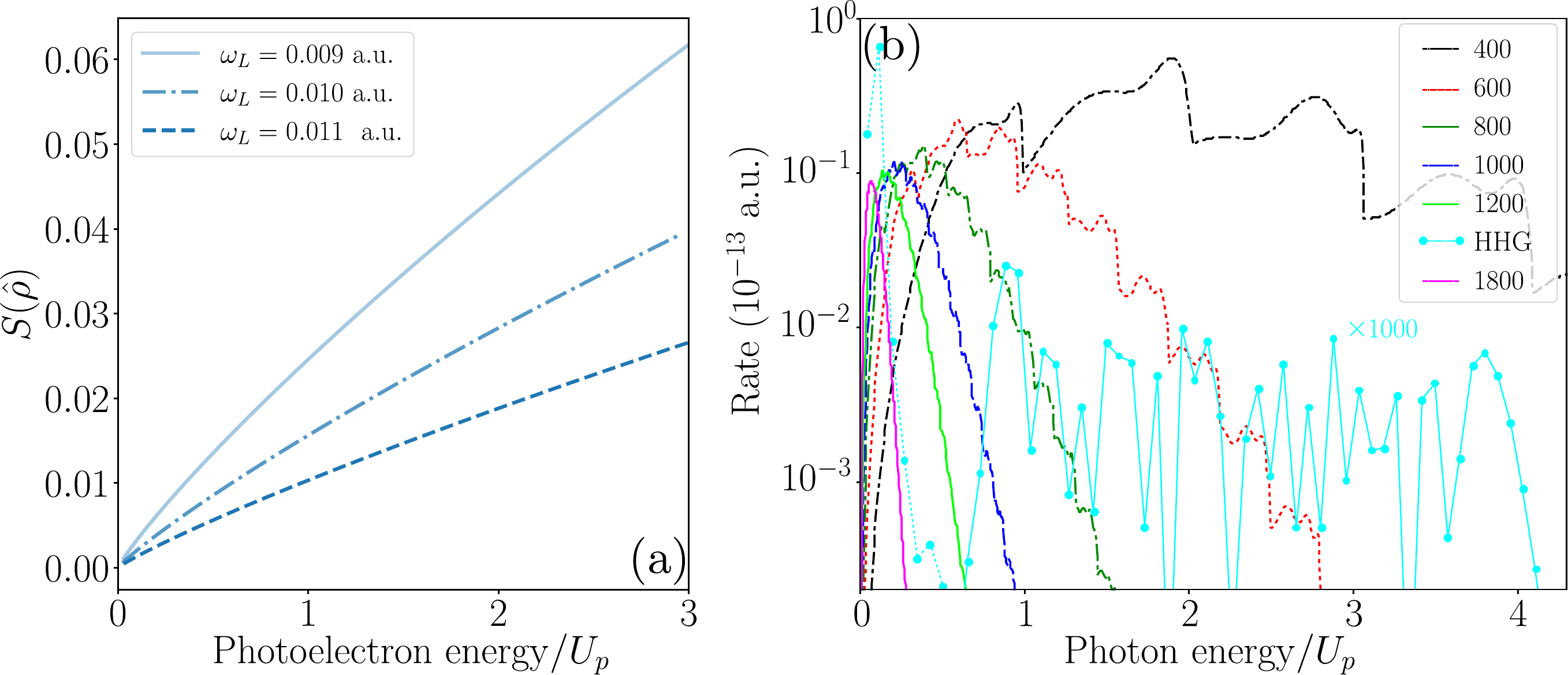}
    \caption{(a) Degree of entanglement between the driving field and photoelectrons emitted along the forward and backward directions for a fixed kinetic energy.~(b) Photon emission probability as a function of frequency for various driving field wavelengths (given in nm in the legend). The cyan curve corresponds to HHG driven by a $\lambda = 1200$ nm field. Each panel adapted from (a) \cite{rivera2022light} and (b) \cite{milovsevic2023quantum}.}
    \label{fig:ATI:class:light}
\end{figure}

For typical strong-field physics driving field conditions ($\lambda_L \approx 800$ nm and $I= 10^{14}$ W/cm$^2$), the values of $\abs{\delta_q}$ remain too small to produce appreciable quantum optical signatures in either the driving fields or the emitted harmonics~\cite{rivera2022light}. In contrast, mid-infrared driving fields give rise to non-negligible entanglement between the driving field and the generated electrons, as illustrated in Fig.~\ref{fig:ATI:class:light}~(a). More specifically, the figure displays the von Neumann entropy associated with the entanglement between electrons of fixed momentum magnitude, propagating in opposite directions and the driving field.

In addition to generating light-matter entanglement, quantum optical treatments of ATI predict the emission of harmonic radiation~\cite{milovsevic2023quantum}. Restricting the attention to continuum electrons with kinetic momentum compatible with direct ATI, the probability of emitting a single photon in the $q$th harmonic is
\begin{equation}
	P(\omega_q)
		= \int \dd \vb{v}\ \abs{\langle 1_q\relaxket{\Phi_{\text{dATI}}(\vb{v},t)}}^2.
\end{equation}
Unlike in HHG, this emission arises from a single-step process. As a consequence, in the single-atom regime the probabilities for the lowest harmonic orders exceed those of HHG by several orders of magnitude under comparable conditions, as shown in Fig.~\ref{fig:ATI:class:light}~(b).~It is, however, an open problem to determine how the emitted radiation scales with the number of emitters and how this scaling is modified by phase-matching conditions.

\subsection{ATI in non-classical driving fields}
\label{sec:ATI_quantum}

Similar to HHG, the prospect of accessing high-intensity squeezed states of light has recently motivated a large number of studies aimed at understanding how non-classical driving fields modify ATI outcomes. These investigations have not only sought to quantify and characterize differences in standard ATI observables when using coherent versus squeezed light~\cite{fang2023strong,wang2023high,lyu_effect_2025,liu_atomic_2025}, but have also explored what additional information about the photon statistics of the driving field can be extracted from the ejected electrons~\cite{heimerl_multiphoton_2024,heimerl_driving_2025,lyu_effect_2025}.~Therefore, this could suggest the use of photoelectron ionization events as an alternative tool for characterizing high-intensity non-classical states of light.

Most of the theoretical approaches build upon the framework originally developed for HHG (see Sec.~\ref{Sec:QHHG:revisited}), with the main distinction lying in the observables under consideration. Unlike HHG, where the focus is on the emitted radiation, ATI primarily probes the matter system, represented in Eq.~\eqref{Eq:state:gen:P} by $\ket{\psi_{\alpha/\beta}(t)}$. Hence, any physical observable $O_{e}$ acting on the electronic sub-space can be evaluated, in the classical limit, as
\begin{equation}\label{Eq:QATI:obs}
	\langle O_e\rangle
		= \int \dd^2 \alpha
			\big[
				\lim_{g\to0}
					P(\alpha,\alpha^*)
			\big]
			\mel{\psi_{\alpha}(t)}{O_e}{\psi_{\alpha}(t)}.
\end{equation}
Examples include the ATI photoelectron spectra~\cite{lyu_effect_2025,wang2023high,heimerl_driving_2025}, where $O_e = \dyad{\vb{p}}$; or the ionization yield in double ionization processes~\cite{liu_atomic_2025}, where $O_e = \int_R \dd\vb{r}_1 \dd \vb{r}_2\dyad{\vb{r}_1,\vb{r}_2}$, with $\vb{r}_1$ and $\vb{r}_2$ denoting the positions of the two electrons involved in the process and $R$ is the spatial region of detection.

\begin{figure}
    \centering
    \includegraphics[width=1\columnwidth]{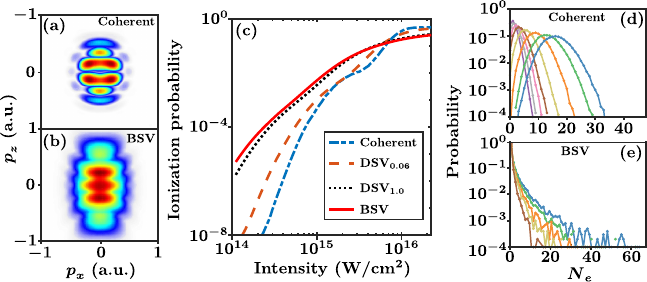}
    \caption{(a),(b) Photoelectron dATI spectra obtained for a coherent state [(a)] and a BSV [(b)] driving field at $\lambda = 500$ nm (adapted from \cite{lyu_effect_2025}). (c) Double-ionization probability for different driving fields, with each curve corresponding to a specific field type. Here, DSV denotes a displaced vacuum state, with the subscript indicating the ratio $I_{\text{coh}}/I_{\text{squ}}$ (adapted from \cite{liu_atomic_2025}). (d), (e) Probability of detecting $N_e$ electrons following a multiphoton ionization of a metal needle tip under a coherent [(d)] and BSV [(e)] field (adapted from \cite{heimerl_multiphoton_2024}).}
    \label{Fig:ATI:Q:Light}
\end{figure}

Similarly to the HHG spectrum, Eq.~\eqref{Eq:QATI:obs} represents an incoherent average of the expectation value of $O_e$, as if the process were driven by an electric field of $E_{\alpha}(t) = \langle \alpha \vert E_Q(t)\vert \alpha \rangle$, with the different realizations of $E_{\alpha}(t)$ weighted according to $\lim_{g\to 0} P(\alpha,\alpha^*)$.~Consequently, typical ATI observables reflect the consequences of this incoherent averaging: a broadening of ATI spectra compared to the coherent state case~\cite{liu_atomic_2025} (displayed in Fig.~\ref{Fig:ATI:Q:Light}~(a) and (b) for dATI), an extension of HATI cutoffs~\cite{wang2023high}, and the erasure of double-knee structures in non-sequential double ionization~\cite{liu_atomic_2025} [Fig.~\ref{Fig:ATI:Q:Light}~(c)].~Intuitively, these effects arise because the different fields $E_{\alpha}(t)$ contribute with distinct ponderomotive potentials $U_p$, which strongly influence both the electron motion in the continuum~\cite{even2024motion} and the resulting momentum interference patterns~\cite{lyu_effect_2025}.~Interestingly, these features were recently confirmed experimentally in strongly-driven metallic needle tips~\cite{heimerl_driving_2025}.~There, post-selection on specific photon numbers in the outgoing BSV driver recovered the expected, classically driven photoelectron spectra, corresponding to the integrand in Eq.~\eqref{Eq:QATI:obs} evaluated at fixed $\alpha$, whereas in the absence of post-selection the spectra exhibited complete blurring, in agreement with the full incoherent integral in Eq.~\eqref{Eq:QATI:obs}.

However, quantum light driven strong-field ionization also highlights the potential of using ionization as a diagnostic tool for the driving field.~At high intensities, direct access to the quantum properties of the driver is experimentally challenging; yet mapping these features onto more accessible systems provides a promising route toward their characterization.~Along these lines, recent experiments on strongly driven metallic needle tips have shown that, in the multiphoton ionization regime, the photon statistics of the driver are imprinted onto the photoelectron spectra~[Fig.~\ref{Fig:ATI:Q:Light}~(c) and (d)], enabling the observation of ionization events that would be essentially inaccessible in the classical regime~\cite{heimerl_multiphoton_2024}.~Building on this idea, recent theoretical work has proposed extending the approach to the tunneling regime~\cite{liu_atomic_2025}.

Finally, while most studies have focused on properties of the emitted photoelectrons, \cite{rivera-dean_microscopic_2025} recently investigated how squeezing modifies the quantum properties of the joint light-matter system compared to the coherent state case [Fig.~\ref{fig:ATI:class:light}~(a)]. It was shown that strong squeezing enhances the effective light-matter coupling, stemming from the relation $S^\dagger(r)aS(r) \approx e^{r}a$ for $r\gg 1$, which in turn substantially increases light-matter entanglement and gives rise to pronounced non-Gaussian features in the driving field when photoelectrons are conditioned onto specific momenta.

\section{Applications to quantum technologies}
\label{sec:application}

\subsection{Application in quantum information processing }

We have seen that the integration of  quantum optical methods in intense light-matter interaction allows to generate non-classical field states, including entangled states or non-classical state superpositions. While much of this success is of fundamental significance for understanding the strong field driven processes such as HHG and ATI, and can be used to leverage typical strong field and attosecond approaches, we also aim to harness these novel quantum states for applications in information processing~\cite{dowling2003quantum, o2009photonic}.
Central to achieve this goal is the experimental verification of the generated non-classical states of light, and the ability to control their properties. However, due to the ultrafast nature of the underlying processes, and the high intensities of the involved light fields, as well as the extreme wavelength ranging towards the XUV regime, traditional measurement schemes reach their technological limitations. 
Therefore, characterization of the generated quantum optical state, as well as the verification of entanglement, ultimately needs novel experimental tools for nonlinear optics, precision metrology and more generally, quantum information science. Only when these are achieved, this field can use its full potential towards applications in quantum information processing. 

\begin{figure}
    \centering
    \includegraphics[width=0.8\columnwidth]{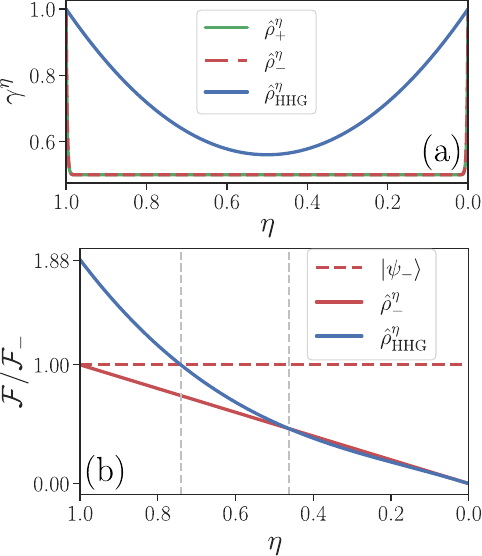}
    \caption{Metrological robustness of the optical cat state generated from HHG, see Eq.~\eqref{Eq:HHG:cond}, under photon loss. (a) The purity $\gamma^\eta$ of the HHG cat state compared to the traditional even ($+$) and odd ($-$) cat states $\ket{\psi_{\pm}} = \ket{\alpha} \pm \ket{- \alpha}$ under photon loss, indicated by the transmission efficiency $\eta$ for input states of an average photon number of $\expval*{N} = 100$. While the purity of the traditional cat states drops dramatically (green and red), the HHG cat state shows enhanced robustness. (b) Quantum Fisher information (QFI) $\mathcal{F}$ for the noisy cat states under photon loss in a phase sensing configuration, normalized to the QFI of the pure odd cat state $\mathcal{F}/\mathcal{F}_-$ in the high photon number regime $\expval*{N} = 100$. The noisy HHG cat state has larger QFI than the pure odd cat state for photon losses up to $\eta = 0.74$ ($25.9 \%$ photon loss), and outperforms the lossy odd cat state up to $53.7\%$ photon loss. The figures have been reproduced from \cite{stammer2024metrological}.}
    \label{fig:metrology}
\end{figure}

Nevertheless, first results on the usefulness of the generated non-classical light states have been proposed. It was predicted that the generated Schrödinger cat states from post-selection schemes in HHG, see Sec.~\ref{sec:conditioning}, show enhanced metrological robustness compared to the traditional cat states~\cite{stammer2024metrological}. While the even ($+$) and odd ($-$) cat states $\ket{\psi_\pm} = \ket{\alpha} \pm \ket{- \alpha}$ are very vulnerable to photon losses~\cite{glancy2008methods}, becoming worse for increasing average photon number, the HHG cat state from Eq.~\eqref{Eq:HHG:cond} is significantly more robust against losses, even for high photon numbers. This is due to the particular structure of the HHG cat state, where the coherent states in the superposition are only distinguishable by $\delta \alpha$, but can have large amplitudes due to $\alpha$. The enhanced robustness therefore comes at the cost of a decreased distinguishability. This can be seen in Fig.~\ref{fig:metrology}~(a), where the purity $\gamma^\eta$ as a function of the transmitivity $\eta$ is shown for all three cat states, indicating that the HHG cat is significantly more robust, even for photon numbers of $\expval*{N}=100$. 
This property further translates to the metrological usefulness by means of the Quantum Fisher information (QFI)~\cite{liu2020quantum}, which provides the precision bound in a parameter estimation scenario. Considering the task of phase estimation, the QFI of the HHG cat state, under photon losses, is larger than the QFI for the noisy even and odd cat states, as shown in Fig.~\ref{fig:metrology}~(b). Strikingly, the QFI of the noisy HHG cat is even larger than the QFI of the pure odd cat state for up to $25.9 \%$ photon loss~\cite{stammer2024metrological}.

While the latest progress for generating non-classical field states using HHG has shown clear approaches to potentially achieve this goal, detecting those states and the corresponding signatures remains a challenge due to the aforementioned extreme properties of the light fields. A crucial, and experimentally important characteristic, is the ultrafast property of the pulses. This raises the natural question if traditional frameworks developed for sufficiently long pulses still work for short non-stationary radiation~\cite{van2025errors, van2025effects}. For instance, it was recently shown that field correlation functions not only depend on the statistics of the field, but also on the shape of the pulse~\cite{van2025errors}. This has crucial consequences since the measured observables, for instance in \cite{theidel2024evidence}, can not be linked directly to the quantum state of the field.

\subsection{Analog simulation of strong field processes}
\label{sec:analog}

A quantum analog simulator is a controllable quantum system engineered to reproduce the dynamics of another, often more complex or less accessible, quantum system, providing an experimental counterpart to purely numerical approaches~\cite{QS1,QS2,QS3}. Instead of attempting to solve intractable equations directly, one lets a well-controlled platform, such as ultracold atoms in tailored optical traps, evolve under designed conditions that mimic the Hamiltonian of the target system. This concept becomes particularly powerful in the realm of attosecond science, where ultrafast electron dynamics driven by intense laser fields unfold on timescales of $10^{-18}-10^{-15}$ s, making them difficult to probe with conventional experiments. In contrast, in an analog simulator the dynamics occur at much slower, experimentally accessible timescales, while still preserving the essential physics of strong-field processes. For example, high-harmonic generation can be simulated in cold atomic gases by applying time-dependent external forces that emulate the action of a strong laser field on electrons. Although neutral atoms do not emit photons, the time-dependent dipole acceleration, a central observable in HHG, can be reconstructed~\cite{arguello2024analog}. This strategy has been shown to reproduce hallmark features of HHG such as the harmonic plateau, the cutoff law, and the suppression of conversion efficiency with increasing ellipticity. Importantly, the absence of collective propagation effects allows clean access to single-particle dynamics, enabling precise benchmarking of theoretical models, while the high tunability of the platform opens pathways to probe parameter regimes that remain inaccessible to direct attosecond experiments. In this sense, quantum analog simulators act as a “magnifying glass” for attosecond science, slowing down ultrafast processes and providing new opportunities to deepen our understanding of strong-field and attosecond phenomena. Furthermore, earlier studies have shown that (i) ultracold atoms in strongly driven optical lattices can emulate electrons in laser-irradiated solids, enabling access to multi-photon and non-perturbative resonances in regimes otherwise difficult to reach with real materials~\cite{Arlinghaus2010}; (ii) trapped Bose–Einstein condensates subjected to time-varying forces reproduce essential features of ultrafast dynamics with temporal magnification of up to twelve orders, allowing direct observation of sub-cycle unbinding and carrier-envelope phase effects~\cite{Senaratne2018}, and more recently that driven Bose-Einstein condensates allow for the generation of non-classical light via HHG~\cite{stammer2025high}; and (iii) proposals for attosecond science simulators based on optically trapped atoms show that fundamental strong-field processes can be faithfully reproduced, while offering the additional advantages of tunable interactions and exotic pulse shapes beyond the reach of conventional lasers~\cite{Sala2017}.

In addition to use analog simulator to investigate the physics of strong field processes, recent advances of integrating the quantum optical approach with analog simulators has been achieved by the quantum Kramers-Henneberger transformation~\cite{arguello2025quantum}. 
The Kramers--Henneberger (KH) transformation describes an atom in a strong oscillating field by moving to the frame of the electron's quiver motion, where the Coulomb potential is shifted as $V_{\text{KH}}(\mathbf{r},t) = V(\mathbf{r}+\alpha(t))$, with $\alpha(t)$ the classical displacement of a free electron in the laser field. In this context, the KH transformation has been extended into the quantum domain by quantizing the displacement motion~\cite{arguello2025quantum}. This quantum KH transformation incorporates the quantum fluctuations of the displacement, yielding quantum electrodynamic corrections absent in the classical picture. The formalism has been developed for both single-mode and continuum-mode oscillators, showing that the effective inertial force acting on the particle can generate squeezing of the oscillator state. As a result, the trapped particle evolves under an effective quantized electric field, with corrections governed by the ratio between the trap’s displacement amplitude and the oscillator’s zero-point motion. Although these corrections are typically perturbative, they become relevant in regimes where the trap’s mechanical degrees of freedom are strongly coupled to the particle. This framework provides a novel pathway towards experimental implementations of quantum optical strong field physics using state-of-the-art optomechanical setups or hybrid cold-atom platforms.

\section{Conclusions: Discussion and Outlook}
\label{conclusions}

\subsection{Discussion}
\label{sec:discussion}

The research field emerging by combining quantum optics and strong-field interactions can be considered as a new frontier in ultrafast science. A novel field where both light and matter are treated quantum mechanically, and where quantum optics and ultrafast science can mutually benefit from each other. This contemporary research field provides new perspectives, in which strong-field processes can be used to generate non-classical states of light and matter, or where quantum noise can influence strong-field driven dynamics.

There are two important aspects of quantum optics of intense light-matter interaction. The first one is fundamental. This new area has already led to numerous theoretical predictions of new phenomena and effects, like generation of the Schr\"odinger cat states, generation of multimode squeezed states, or the generation of harmonics by bright squeezed states. Moreover, some of these theoretical predictions have been observed in experiments. The second aspect is more applied and technologically oriented. Quantum optics of intense light--matter interaction develops a new kind of quantum spectroscopy on the ultrafast, attosecond time-scales. This novel spectroscopy incorporates and employs information about the quantum nature of electromagnetic fields. 

\subsection{Future directions}
\label{sec:outlook}

There are several directions we foreshadow for this young field for further development~\cite{Ciappina2025JOpt}. First, how can strong field physics provide new quantum information technologies, and second, how can quantum information enhance measurements in strong field physics? In the first case, here we list four evident challenges for the related research field. They are all based on the necessary theoretical ingredients: development of quantum field-theoretical treatments of strong field physics \cite{stammer2023quantum}, but they all contribute toward new quantum information applications. In the second case, on one hand we encourage the old Greek saying: \textit{Cobbler, stick to your last}. This is to emphasize that one should not dismiss traditional strong field settings and measurements, which have shown tremendous success over the past decades. We rather encourage the use of quantum optical effects to leverage those measurements, e.g. by increasing measurement sensitivities due to reduced field fluctuations. On the other hand, we will seek new measurement and detection methods based on quantum information science to enhance the power of the new approach for generating non-classical states of light: 

\begin{itemize}
 
\item{\bf Challenge 1: Generation of massively quantum states of light}. New systematic methods should be established to produce large-photon-number entangled states, extending conditioning and post-selection techniques from atoms and molecules to correlated solids~\cite{bhattacharya2023strong, lamprou2024generation}.~Generation of multimode squeezed and more exotic quantum states will be possible, going beyond the negligible depletion/excitation limit~\cite{stammer2024entanglement}, or using HHG in resonant media~\cite{Misha2025resonant}. Alternatively, one will use bright squeezed light~\cite{gorlach2023high,rasputnyi2024high}, or its combination with conventional intense laser pulses~\cite{lemieux2024photon}, to generate high harmonics. Structured light will be used to study geometrical, chiral, and topological states in QED of attoscience, potentially enabling to probe and control many-body quantum systems at an unprecedented scale.

\item{\bf Challenge 2: Generation of massively quantum states of light and matter}. One will explore entanglement between quantized light fields and electronic states in complex materials, focusing on ultrafast processes such as ATI, HHG, and other multi-electron processes. This will provide the first systematic framework for observing and exploiting light-matter entangled states in solids, using similar approaches as mentioned above. Earlier work has shown that conditioning on ATI events or on distinct HHG recombination paths in molecules and simple solids can generate light-matter entanglement. Recent advances~\cite{laurell2025measuring} concerning the reconstruction of the photoelectron density matrix provide a new testing ground. These studies have revealed how classical and quantum noise reduce purity, but the open question remains: how does photon entanglement influence the reconstructed density matrix, and can experimental data unambiguously signal the presence of underlying quantum correlations? This connects QED of attoscience to the broader framework of multi-fragment “Zerfall” processes.

\item{\bf Challenge 3: Characterization and exploitation of the generated states}. One will have to develop ultrafast quantum-optical methods to verify and quantify entanglement, providing new tools for nonlinear optics, precision metrology, and quantum information. This will open pathways to applying attosecond-scale entanglement across multiple fields. For the first steps toward applications in nonlinear optics and metrology, see~\cite{lamprou2025nonlinear,stammer2024metrological}.

\item{\bf Challenge 4: Simulation of QED of attoscience with quantum platforms}. One may be able to design and build quantum simulators based on cold atoms and ions, enabling controlled studies of strong-field dynamics such as quantum trap shaking~\cite{arguello2024analog}, and generalized Kramers-Henneberger transformations~\cite{arguello2025quantum}. This will establish a new route to test and extend attoscience concepts under fully tunable conditions.
\end{itemize}

\begin{acknowledgments}

P.S. acknowledges funding from the European Union’s Horizon 2020 research and innovation program under the Marie Skłodowska-Curie grant agreement No 847517. P.T. acknowledges the Hellenic Foundation for Research and Innovation (HFRI) and the General Secretariat for Research and Technology (GSRT) under grant agreement CO2toO2 Nr.:015922, the European Union’s HORIZON-MSCA-2023-DN-01 project QU-ATTO under the Marie Skłodowska-Curie grant agreement No 101168628 and ELI--ALPS. ELI--ALPS is supported by the EU and co-financed by the European Regional Development Fund (GINOP No. 2.3.6-15-2015-00001).
M.~F.~C.~acknowledges support by the National Key Research and Development Program of China (Grant No.~2023YFA1407100), Guangdong Province Science and Technology Major Project (Future functional materials under extreme conditions - 2021B0301030005), the Guangdong Natural Science Foundation (General Program project No. 2023A1515010871), and the National Natural Science Foundation of China (Grant No. 12574092).
ICFO-QOT group acknowledges support from:
European Research Council AdG NOQIA; MCIN/AEI (PGC2018-0910.13039/501100011033,  CEX2019-000910-S/10.13039/501100011033, Plan National STAMEENA PID2022-139099NB, , project funded MCIN and  by the “European Union NextGenerationEU/PRTR" (PRTR-C17.I1), FPI); QUANTERA DYNAMITE PCI2022-132919, QuantERA II Program co-funded by European Union’s Horizon 2020 program under Grant Agreement No 101017733; Ministry for Digital Transformation and of Civil Service of the Spanish Government through the QUANTUM ENIA project call - Quantum Spain project, and by the European Union through the Recovery, Transformation and Resilience Plan - NextGenerationEU within the framework of the Digital Spain 2026 Agenda; MICIU/AEI/10.13039/501100011033 and EU (PCI2025-163167); Fundació Cellex;  Fundació Mir-Puig; Generalitat de Catalunya (European Social Fund FEDER and CERCA program; Barcelona Supercomputing Center MareNostrum (FI-2023-3-0024); 
Funded by the European Union (HORIZON-CL4-2022-QUANTUM-02-SGA, PASQuanS2.1, 101113690, EU Horizon 2020 FET-OPEN OPTOlogic, Grant No 899794, QU-ATTO, 101168628),  EU Horizon Europe Program (No 101080086 NeQSTGrant Agreement 101080086 — NeQST).

\end{acknowledgments}

\bibliography{literature}

\end{document}